\documentclass[%
preprint,
 amsmath,amssymb,
prc,
]{revtex4-2}

\usepackage{relsize}
\usepackage{physics}
\usepackage{graphicx}% Include figure files
\usepackage{dcolumn}% Align table columns on decimal point
\usepackage{bm}% bold math
\begin{document}

\preprint{APS/123-QED}

\title{P-even and -odd asymmetries on $^{117}$Sn at the vicinity of the p-resonance E$_\mathrm{p}$=1.33 eV}

\author{L. E. Char\'on-Garc\'ia}
\email{lcharon@estudiantes.fisica.unam.mx}
\affiliation{
 Instituto de F\'isica, Universidad Nacional Aut\'onoma de M\'exico, Apartado Postal 20-364, 01000, M\'exico 
}

\author{J. Curole}
\email{jcurol4@gmail.com}
\affiliation{
 Indiana University, Bloomington, IN, 47405, USA
}

\author{V. Gudkov}
\email{gudkov@sc.edu}
\affiliation{
 University of South Carolina, Columbia, South Carolina 29208, USA 
}

\author{L. Barr\'on-Palos}
\email{libertad@fisica.unam.mx}
\affiliation{
 Instituto de F\'isica, Universidad Nacional Aut\'onoma de M\'exico, Apartado Postal 20-364, 01000, M\'exico 
}

\author{W. M. Snow}
\email{wsnow@indiana.edu}
\affiliation{
 Indiana University, Bloomington, IN, 47405, USA 
}
\collaboration{NOPTREX Collaboration}

\date{\today}% It is always \today, today,
             %  but any date may be explicitly specified

\begin{abstract}
A self consistent description of  angular correlations in neutron induced reactions is required for quantitative analysis of parity violating (PV) and time reversal invariance violating (TRIV) effects in neutron nucleus scattering.
 The 1.33 eV p-wave compound resonance in $^{117}$Sn is one of the few p-wave resonances where enough measurements have been performed to allow a nontrivial test of the internal consistency of the theory. We present the results of a global analysis of the several different asymmetries and angular distribution measurements in ($n, \gamma$) reactions on the 1.33 eV p-wave resonance in $^{117}$Sn conducted over the last few decades. We show  that the compound resonance mixing theory can give an internally consistent description of all observations made in this system to date within the experimental measurement errors. We also confirm the conclusions of previous analyses that a subthreshold resonance in $^{117}$Sn dominates correlations related to s-p mixing, and  discuss the implications of these results for future searches for TRIV in this system.
\end{abstract}

%\keywords{Suggested keywords}%Use showkeys class option if keyword
                              %display desired
\maketitle

%\tableofcontents

\section{\label{sec:introduction}Introduction}

The angular distributions in ($n, \gamma$) reactions on the 1.33 eV p-wave resonance in $^{117}$Sn measured in different experiments over the last few decades have been a subject of discussions due to reported inconsistencies in descriptions of experimental data.  In 1984, Alfimenkov \textit{et al.} \cite{Alfimenkov1984} measured the left-right asymmetry  in radiative neutron capture at the vicinity of $^{117}$Sn 1.33 eV p-wave resonance.  One year later this group measured the ($n, \gamma$)  forward-backward asymmetry \cite{Alfimenkov1985} at the same resonance. However, the analysis of these data could not consistently describe the experimental results using theoretical descriptions of ($n, \gamma$) correlations \cite{Flambaum:1983ek} (see for details ~\cite{alfimenkov1991p}).
This analysis was revisited by other authors~\cite{Sharapov, Barabanov}, but they also could not consistently describe all observable parameters. It should be noted that in the above analyses a two-level approximation was used with a p-wave resonance at 1.33 eV and a negative s-wave resonance at \mbox{-29.2} eV  (the importance of multi resonance analysis  was  pointed out in~\cite{Gudkov1991}).
 We show  that a resonance description of neutron radiative capture \cite{Flambaum:1983ek} with three resonances can give an internally consistent description of the experimental data reerred to above. This not only resolves this particular problem for the interpretation of angular correlations in low energy neutron radiative capture, but gives an assurance in the measurements of nuclear spectroscopic parameters required for new experiments in a search for time reversal invariance violation (TRIV) in neutron scattering.

The search for new sources of TRIV is one of the highest  priorities in nuclear, particle and astrophysics. New sources of CP/T violation beyond the Standard Model are needed to explain the observed matter-antimatter asymmetry of the universe. Theoretical progress to identify the large number of possible sources for T violation has made it very clear that any single type of T violation search cannot be equally sensitive to all possible mechanisms. Therefore it is important to pursue possible experiments in different systems which can provide sufficient sensitivity to discover something new in physics or to improve the current limits.

One of the possible experiments with a high sensitivity to TRIV is related to neutron scattering at p-wave compound resonances,  where  TRIV effects can be observed as asymmetry in polarized neutron transmission through polarized nuclear targets. Multiple papers \cite{Gudkov:2013nwa,PhysRevC.90.065503,Bunakov:1982is,GUDKOV199277,BedaSkoy2007} have suggested that TRIV searches in complex nuclei have the potential to  improve the current experimental limit from electric dipole moment (EDM) experiments for   parity-odd (P-odd) and time-reversal -odd (T-odd) interactions beyond the Standard Model (BSM). Also, neutron scattering experiments  on  complex nuclei can improve on the existing experimental upper bounds on P-even/T-odd BSM interactions by about three orders of magnitude~\cite{Gudkov:1991qc,Song:2011jh}. The opportunity to have different nuclear systems to search for TRIV  helps in providing assurance that possible ``accidental" cancellation of T violating effects due to unknown structural factors related to the strong interactions in particular systems can be avoided. It may also possess different sensitivity to many possible  new sources of T violation BSM~\cite{Pospelov:2001ys,POSPELOV2005119,PhysRevC.84.025501,PhysRevC.87.015501}.

The basic approach for TRIV experiments  in neutron scattering on heavy nuclei involves the measurement  of T-odd correlations in the neutron forward scattering, such as $\vec{{\sigma}}{_n} \cdot (\vec{k}{_n} \times \vec{I})$ where $\vec{\sigma}{_n}$ is the spin of the neutron, $\vec{k}{_n}$ is the neutron momentum, and $\vec{I}$ is the spin of the nucleus. This correlation violates   P- and T- invariance and is free from  so-called final state effects, which in other processes may mimic TRIV~\cite{Gudkov:2013nwa,PhysRevC.90.065503,GUDKOV199277}. These TRIV effects are directly related to the P violating effects by a simple spin dependent factor~\cite{PhysRevC.97.065502}. P violating effects in n-A resonances in several heavy nuclei were measured by the TRIPLE collaboration at LANSCE a couple of decades ago, following earlier work and discoveries of large amplifications of P-odd effects on p-wave neutron-nucleus resonances at other laboratories (see, for example~\cite{Mitchell2001157} and references therein). The dozens of measurements of P-odd asymmetries through the neutron helicity dependence of the cross section were analyzed and interpreted in terms of the mixing of opposite parity components of s- and p-resonances through the parity-odd components of the electroweak interaction. For the great majority of the p-wave resonances where large parity-odd asymmetries were measured, however, there is no other experimental data available on other types of angular correlations beyond the P-odd  cross section helicity dependence and the p-wave resonance energy and total width. The theory of resonance mixing has implications for both P-even and P-odd observables. Therefore, to determine additional parameters of the resonances, it is useful to expand the range of observables under analysis to include P-even angular correlations and angular correlations in neutron radiative capture.

The 1.33 eV p-wave resonance in the $^{117}$Sn nucleus is one of the most interesting candidates for a time reversal test in polarized neutron transmission through a polarized nuclear target, because a rather large P-odd effect has been measured at this resonance. Also, the $I=1/2$ spin of $^{117}$Sn is a great advantage for TRIV test because spin $1/2$ systems do not contain admixture of higher degrees tensor polarizations, which can lead to systematic errors in neutron spin interactions and make the buildup of large ensembles of highly-polarized $^{117}$Sn nuclei more difficult due to spin relaxation from interaction with electric field gradients. A large value of polarization of $^{117}$Sn nuclei can be produced by a new technique   called SABRE~\cite{sabre1,sabre2,C6CC07109K}. SABRE involves transfer of the spin order from parahydrogen molecules to another pair of atoms in groups of four atoms with asymmetric spin couplings as the molecules are transiently adsorbed onto certain molecular catalysts. In principle this process can maintain a large steady-state $^{117}$Sn polarization over the timescales needed for a TRIV search.

\newpage

\section{\label{sec:introduction}P-even effects in (n,$\gamma$) reactions}

In 1984 Flambaum and Sushkov developed a theoretical formalism for the description of P-odd and P-even angular correlations in neutron radiative capture at very low neutron energies~\cite{Flambaum:1983ek,Barabanov}. For an unpolarized target and in experiments where the gamma polarization is not measured, this expression becomes

\begin{equation}
\begin{split}
\frac{d\sigma}{d\Omega}&= \frac{1}{2}\left[a_{0}+a_{1}\vec{n}_{n}\cdot\vec{n}_{\gamma}+a_{2}\vec{s}_{n}\cdot\left[\vec{n}_{n}\times\vec{n}_{\gamma}\right]+a_{3}\left(\left(\vec{n}_{n}\cdot\vec{n}_{\gamma}\right)^{2}-\frac{1}{3}\right)\right] \\
& =a_{0}+a_{1}\,cos \,(\theta)+a_{2}f_{n}sin\, (\theta)\, cos\, (\phi) + a_{3}\left(cos^{2}\,(\theta)-\frac{1}{3}\right),
\end{split}
\label{eq:equation1}
\end{equation}

where $\vec{n}_{n(\gamma)}$ and $\vec{s}_{n}$ are unit vectors along the neutron beam ($\gamma$ emission) and neutron spin direction, respectively, and $f_{n}$ is the neutron polarization. 

Thus for $^{117}$Sn, one can define  $k$ as the neutron wave-vector, $E$ as the neutron energy, $E_{s(p)}$ as the energy of the s(p)-resonance, $\Gamma_{s(p)}$ as the total width of the s(p)-resonance, and introducing  the amplitude of the capture channel of neutrons $\gamma^{n}_{s(p)}=\sqrt{\Gamma^{n}_{s(p)}}$   with $l=0(1)$ or $j=1/2(3/2)$, and $\gamma^{\gamma}_{s(p)}=\sqrt{\Gamma^{\gamma}_{s(p)}}$ is the amplitude of the gamma decay channel of the compound s(p)-state. It should be noted that the sign of the neutron width amplitude is also an important spectroscopic parameter which can be obtained only from interference terms in angular correlations. At present, there are no experimental results which determine these signs for the low-lying resonances in $^{117}$Sn.  In the three-level approximation case (one p- and two s-resonances), we have an interference term between s-resonances in $a_{0}$ of the form \cite{Flambaum:1983ek}

\begin{equation}
\frac{\gamma^{n}_{s1}\gamma^{n}_{s2}
    \gamma^{\gamma}_{s1}\gamma^{\gamma}_{s2}\left((E-E_{s1})(E-E_{s2})+\Gamma_{s1}\Gamma_{s2}/4\right)}{\left[(E-E_{s1})^{2}+\frac{\Gamma^{2}_{s1}}{4}\right]\left[(E-E_{s2})^{2}+\frac{\Gamma^{2}_{s2}}{4}\right]}.    \label{eq:equation8}
\end{equation}
\vspace*{-0.5cm}

Since the sign of every single amplitude in Eq.(\ref{eq:equation8}) is an important unknown parameter, the interference term could be positive (constructive interference) or negative (destructive interference). For nuclei with spin $I \neq 0$ ($I=1/2$ for $^{117}$Sn) and in the neutron total angular momentum representation $j$, the total gamma width has two contributions from $j=1/2$ and $j=3/2$ p-wave neutron capture channels \cite{Flambaum:1983ek}
\vspace*{-0.5cm}

\begin{equation}
    \Gamma^{n}_{p}=\Gamma^{n}_{p1/2}+\Gamma^{n}_{p3/2}.
        \label{eq:equation9}
\end{equation}

At present, there are no separately-measured experimental values for $\Gamma^{n}_{p1/2}$ and $\Gamma^{n}_{p3/2}$ for $^{117}$Sn  at the p-resonance $E_{p}=1.33$ eV. Eqs. (2) and (3) can be expressed as functions of the relative amplitudes

\begin{equation}
x^{2} =\frac{\Gamma^{n}_{p1/2}}{\Gamma^{n}_{p}} \qquad \mathrm{and } \qquad y^{2} =\frac{\Gamma^{n}_{p3/2}}{\Gamma^{n}_{p}}.
\label{eq:equation10}
\end{equation}

Since $x^{2}+y^{2}=1$ we can take

\begin{equation}
    x = cos (\phi) \qquad \mathrm{and} \qquad  y = sin (\phi).
\label{eq:equation11}
\end{equation}

Past efforts to obtain the signs and values for $x$ and $y$ measuring the P-even asymmetries in (n,$\gamma$) reactions using the left-right asymmetry  $\epsilon^{L-R}$,  the forward-backward asymmetry $\epsilon^{F-B}$, and p-wave angular anisotropy $\epsilon^{a}_{p}$ \cite{Alfimenkov1985,Sharapov,Barabanov} were not successful, where:

\begin{equation}
\begin{split}
    \epsilon^{L-R}(E)&=\frac{\sigma(90^{0},0^{0},E)-\sigma(90^{0},180^{0},E)}{f_{n}[\sigma(90^{0},0^{0},E)+\sigma(90^{0},180^{0},E)]} \\
    & =\frac{a_{2}}{a_{0}-a_{3}/3},
\end{split}
\label{eq:equation12}
\end{equation}

\begin{equation}
\begin{split}
    \epsilon^{F-B}(E,\theta=45^{0})=&\frac{\sigma(\theta=45^{0},E)-\sigma(\theta=135^{0},E)}{\sigma(\theta=45^{0},E)+\sigma(\theta=135^{0},E)}\\ & =\frac{1}{\sqrt{2}}\frac{a_{1}}{a_{0}+a_{3}/6},
\end{split}
\label{eq:equation13}
\end{equation}

\begin{equation}
\begin{split}
\epsilon^{a}(\theta)&=\frac{2\sigma_{p}(90^{0},E_{p})}{\sigma_{p}(\theta,E_{p})+\sigma_{p}(\pi-\theta,E_{p})} \\
&= \frac{U^{2}_{p}-(a_{3}/3)}{U^{2}_{p}+(a_{3}/3)\left(3\, cos^{2}(\theta)-1\right)}.
\end{split}
\label{eq:equation14}
\end{equation}

From one of these observables we can obtain a pair of absolute values $|x|$ and $|y|$, but not the signs. The correct determination of both the absolute values and the signs should allow one to reproduce experimental values for these three P-even asymmetries. On the other hand, the spectroscopic parameter

\begin{equation}
\begin{split}
    t^{2}_{\theta}(E_{p})&= \frac{d\sigma_{p}(\theta,E_{p})/d\Omega}{d\sigma_{s}(\theta,E_{p})/d\Omega} \\
    &=\frac{U^{2}_{p}(E_{p})}{U^{2}_{s}(E_{p})}\left[1+\alpha\left(cos^{2}(\theta)-\frac{1}{3}\right)\right],
\end{split}
\label{eq:equation15}
 \end{equation}

with $\alpha =-(3/\sqrt{2})xy-(3/4)y^{2}$ characterizes the relative contribution to the total cross section from both the s- and p-resonances measured at the p-resonance energy $\sigma_{p}(\theta,E_{p})/\sigma_{s}(\theta,E_{p})$ \cite{Barabanov,Sharapov}. It can be shown that

\begin{equation}
    t^{2}(55^{0},E_{p})=\frac{U^{2}_{p}(E_{p})}{U^{2}_{s}(E_{p})}=\frac{\sigma^{\gamma}_{p}(E_{p})}{\sigma^{\gamma}_{s}(E_{p})}
    \label{eq:equation16}
\end{equation}

and

\begin{equation}
    \epsilon^{a}_{p}(\theta)
    =\frac{t^{2}(90^{0},E_{p})}{t^{2}(\theta,E_{p})}.
    \label{eq:equation17}
\end{equation}

The knowledge of $\phi$ in Eq. (\ref{eq:equation11}) allows to evaluate the parameter $\kappa(J)$, which relates
 the P-odd difference of total cross sections $\Delta\sigma_{P}$ proportional to the correlation $\vec{\sigma}_{n}\cdot \vec{k}_{n}$, and the TRIV ones $\Delta\sigma_{PT}$ which are proportional to $\vec{\sigma}_{n}\cdot[\vec{k}_{n}\times\vec{I}]$ in neutron transmission experiments to the corresponding nuclear matrix elements $v$ and $w$ \cite{PhysRevC.97.065502}

\begin{equation}
 \frac{\Delta\sigma_{PT}}{\Delta\sigma_{P}}=\kappa(J) \frac{w}{v},
\label{eq:deltaPT-deltaP}
\end{equation}

In $^{117}$Sn ($I=1/2$) case

\begin{equation}
    \kappa(J)=\left[\frac{\sqrt{I}}{2(I+1)}\right]\frac{(-2\sqrt{I}\,x+\sqrt{2I+3}\,y)}{x} \qquad J=I+\frac{1}{2},
    \label{eq:kappa_J}
\end{equation}

for the coupling scheme $\vec{\hat{J}}=\vec{\hat{I}}+(\vec{\hat{l}}+\vec{\hat{s}})$  used in \cite{Flambaum:1983ek}.

This parameter is a function of the partial neutron width amplitudes $x$ and $y$, and thus a function of $\phi$. From  \ref{eq:kappa_J}, in case $\kappa(J)=0$ the PT-odd effect is cancelled regardless the magnitude of $w$. Since each p-resonance has a  specific value of $\kappa(J)$, for some resonances it might be large, which would make a future experiment on such a resonance more sensitive to the TRIV amplitude of interest.  Therefore, before the designing  TRIV experiment it is important to find a good nuclear target candidate with both a large P-odd asymmetry and with a large value of the parameter $\kappa(J)$. With this objective it is important to analyze  previous experiments to characterize the low-lying p- and s-resonances with the objective to determine the value of $\phi$.

\section{Experimental values for spectroscopic parameters and P-even effects for $^{117}$Sn}

The nearest compound resonances to $E_{p}=1.33$ eV p-wave resonance in $^{117}$Sn are a positive s-resonance at 38.8 eV and a sub-threshold state (a negative s-resonance)  \cite{Mughabghab1981,Mughabghab2006}. The determination of the spectroscopic parameters of this negative s-resonance is difficult because we cannot directly measure the properties of this state: it can only be inferred from information on the neutron energy dependence of the cross section near zero energy. Furthermore, we find different reported values for the resonance energy of this sub-threshold state in different global data evaluations: -29.2 eV (Table \ref{tab:table1} and Ref. \cite{Mughabghab1981}) and -81.02 eV (Ref. \cite{Mughabghab2006}). The change of the position of the subthreshold resonance from -29.2 eV to -81.02 eV in the 2006 re-evaluation~\cite{Mughabghab1981,etde_20332542} occurred despite the fact that the spectroscopic values for positive resonances had either not changed at all or changed only very slightly. We believe that this shift comes from small changes in how the data fitting procedure was reoptimized to determine some global parameters of interest in nuclear data evaluations such as level densities, Westcott g-factors, and resonance integrals parameters. It is known that in these global evaluations the negative energy resonance parameters often serve the function of ``absorbing" some of these changes. Therefore this later change in the inferred value for the subthreshold resonance position in $^{117}$Sn appearing in the global data evaluations may have nothing to do with the real properties of the resonance. We therefore choose to rely on data taken close to zero neutron energy, where the effects of the tail of the subthreshold resonance are the largest. Alfimenkov, Lyapin \textit{et al.} measured P-even effects and spectroscopic parameters for these resonances in $^{117}$Sn (Tables and Figs. \ref{tab:table1}-\ref{tab:table2}) \cite{doi:10.1080/00337578608208388,ALFIMENKOV198393,Alfimenkov1984,alfimenkov1984gamma,Barabanov,Sharapov,Smith2001} in the low neutron energy regime. 

\begin{table}
\caption{Spectrometric parameters for $^{117}$Sn$+n$ by Alfimenkov (refs. \cite{ALFIMENKOV198393} and \cite{alfimenkov1984gamma}) and Mughaghab (Ref. \cite{Mughabghab1981})} 
\begin{tabular}{ccccc}
\hline
\hline
$E_{s}$ (eV) &  $\Gamma^{n0}_{s}$ (meV)  &   $\Gamma_{s}$ (meV)  &    $\Gamma^{\gamma}_{s}$ (meV) \\ 
\hline 
-29  & 7.33   & 100   & 2.3 $\pm$ 0.4\\
38.80 $\pm$ 0.05   &  0.67 $\pm$ 0.02  & 100  & 0.6 $\pm$ 0.2 \\
\hline
 $E_{p}$ (eV) &  $\Gamma^{n}_{p}$ (meV) &  $\Gamma_{p}$ (eV) &  $\Gamma^{\gamma}_{p}$ (meV)  \\ 
\hline
1.33 $\pm$ 0.01  & (2.5 $\pm$ 0.2)$\times 10^{-4}$  &  0.148 $\pm$ 0.01  & 1.2 $\pm$ 0.3  \\
\hline
\hline
\end{tabular}
\label{tab:table1}
\end{table}

\begin{table}
\caption{Values for spectrometric parameters $\epsilon^{a}_{p}$ and $t^{2}_{\theta}(E_{p})$ for $^{117}$Sn$+n$ by Alfimenkov (Ref. \cite{doi:10.1080/00337578608208388}) and Lyapin (Ref. \cite{Lyapin}).}
\begin{tabular}{ccccc}
\hline 
\hline
\multicolumn{1}{c}{$t^{2}(\theta,E_{p})$}  & $\theta$ & $\epsilon^{a}_{p}(\theta)$ & $\theta$ \\
\hline 
1.83 $\pm$ 0.18  & $45^{0}$ & 1.18 $\pm$ 0.12 & $45^{0}$   \\ 
3.0 $\pm$ 0.3 & $90^{0}$    & 1.63 $\pm$ 0.14 & $45^{0}$   \\ 2.16 $\pm$ 0.22 & $90^{0}$ & &
\\
\hline
\hline
\end{tabular}
\label{tab:table2}
\end{table}

\begin{figure}
\includegraphics[width=3.2in]{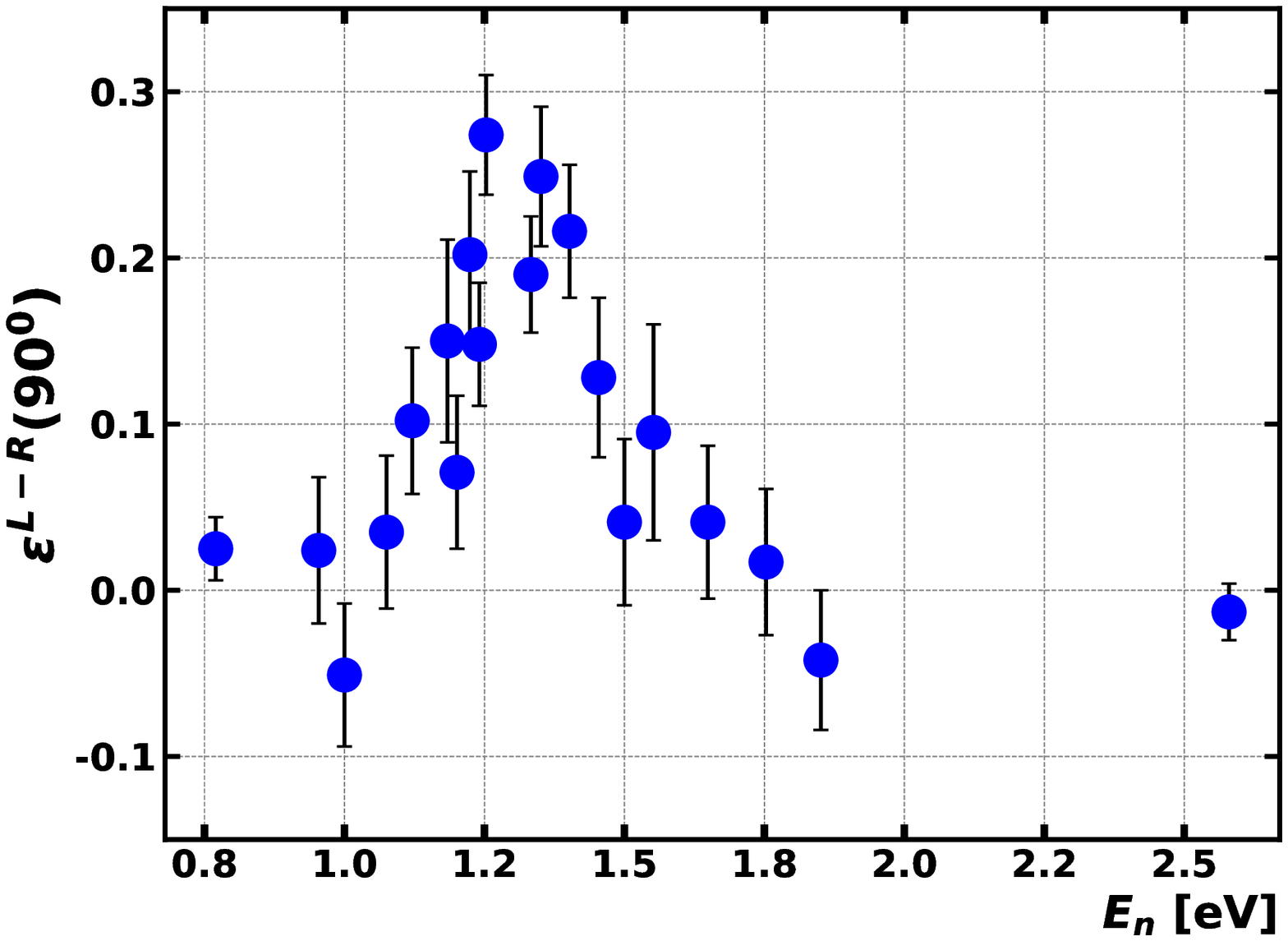}
\includegraphics[width=3.2in]{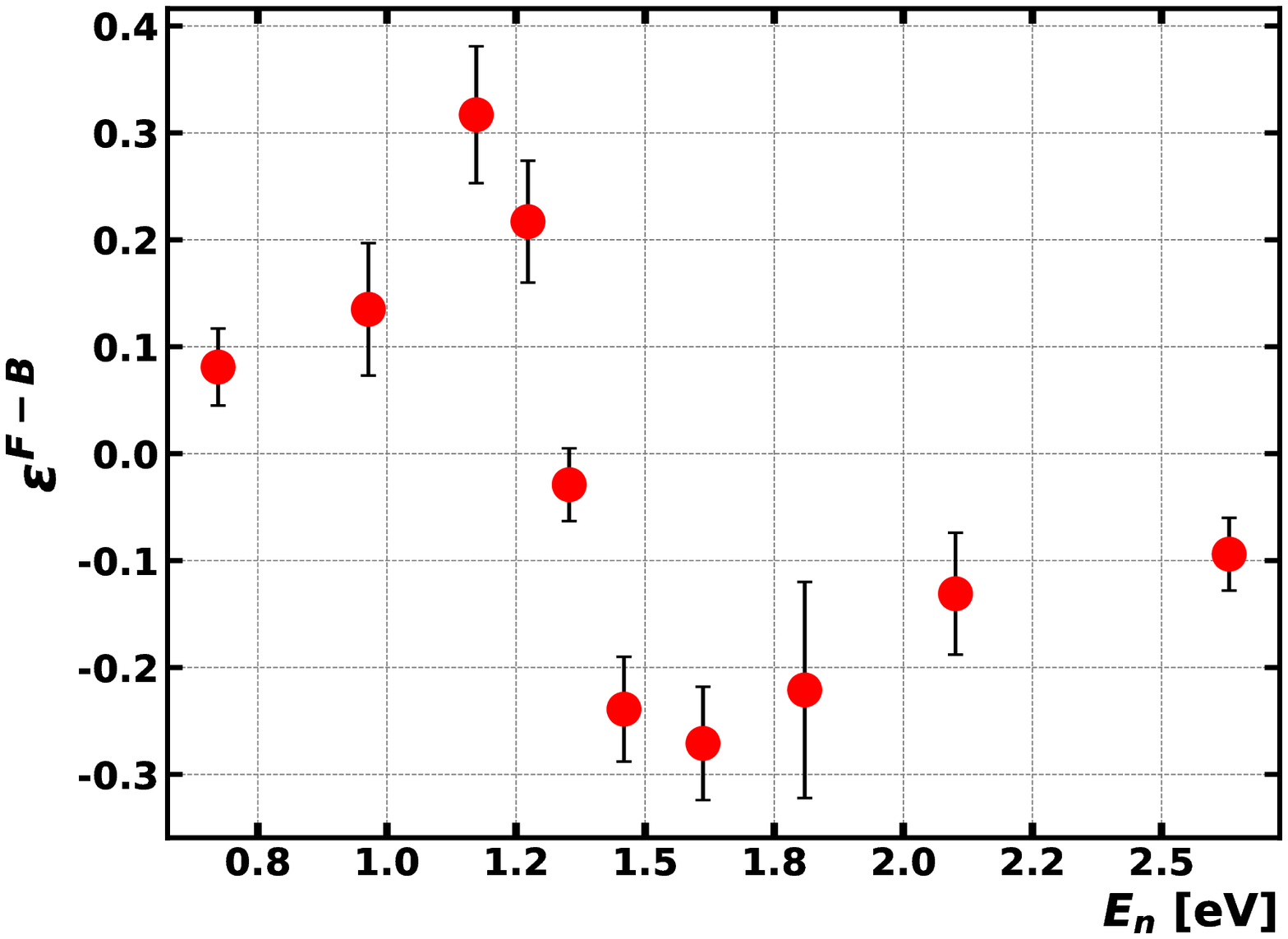}
\caption{Results of the asymmetry $\epsilon^{L-R}$ (left figure from refs. \cite{Alfimenkov1984,Sharapov}) and $\epsilon^{F-B}$ (right figure from refs.~\cite{Alfimenkov1985,Sharapov}) for $^{117}$Sn at the vicinity of the \textit{p}-resonance $E_{p}=1.33$ eV.} 
\label{fig:figure1}
 \end{figure}

In this paper we conduct an analysis to seek consistency between the theoretical formalism \cite{Flambaum:1983ek} and experimental results for these P-even asymmetries using the spectroscopic parameters measured by Alfimenkov, the -29.2 eV negative s-resonance by Mughabghab~\cite{Mughabghab1981}, later measurements of $^{117}$Sn resonance parameters by the TRIPLE collaboration~\cite{Smith2001}, and a recent measurement of an angular-dependence of the (n, $\gamma$) p-wave resonance shape~\cite{Koga2022}. Then, the self consistent description gives a unique $\phi$ value by reproducing experimental values for all these observables.

\section{Seeking for consistency between Flambaum-Sushkov formalism and measured P-even correlations in $^{117}$Sn$+n$.}

\subsection{Left-right $\epsilon^{L-R}$ and forward-backward $\epsilon^{F-B}$ asymmetries}

\subsubsection{Two-level approximation}

\begin{figure}[h!]
\includegraphics[width=3.2in]{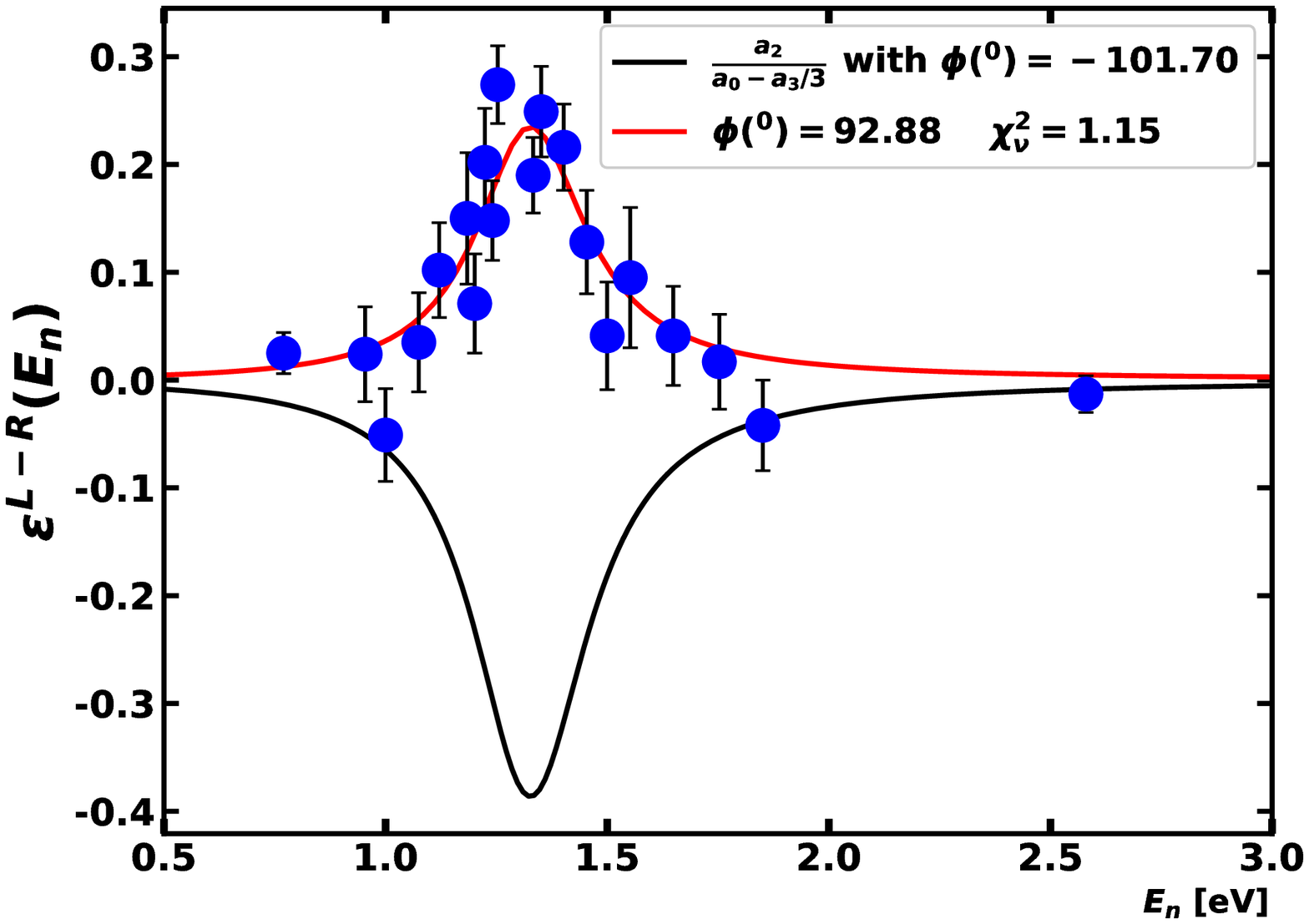}
\includegraphics[width=3.2in]{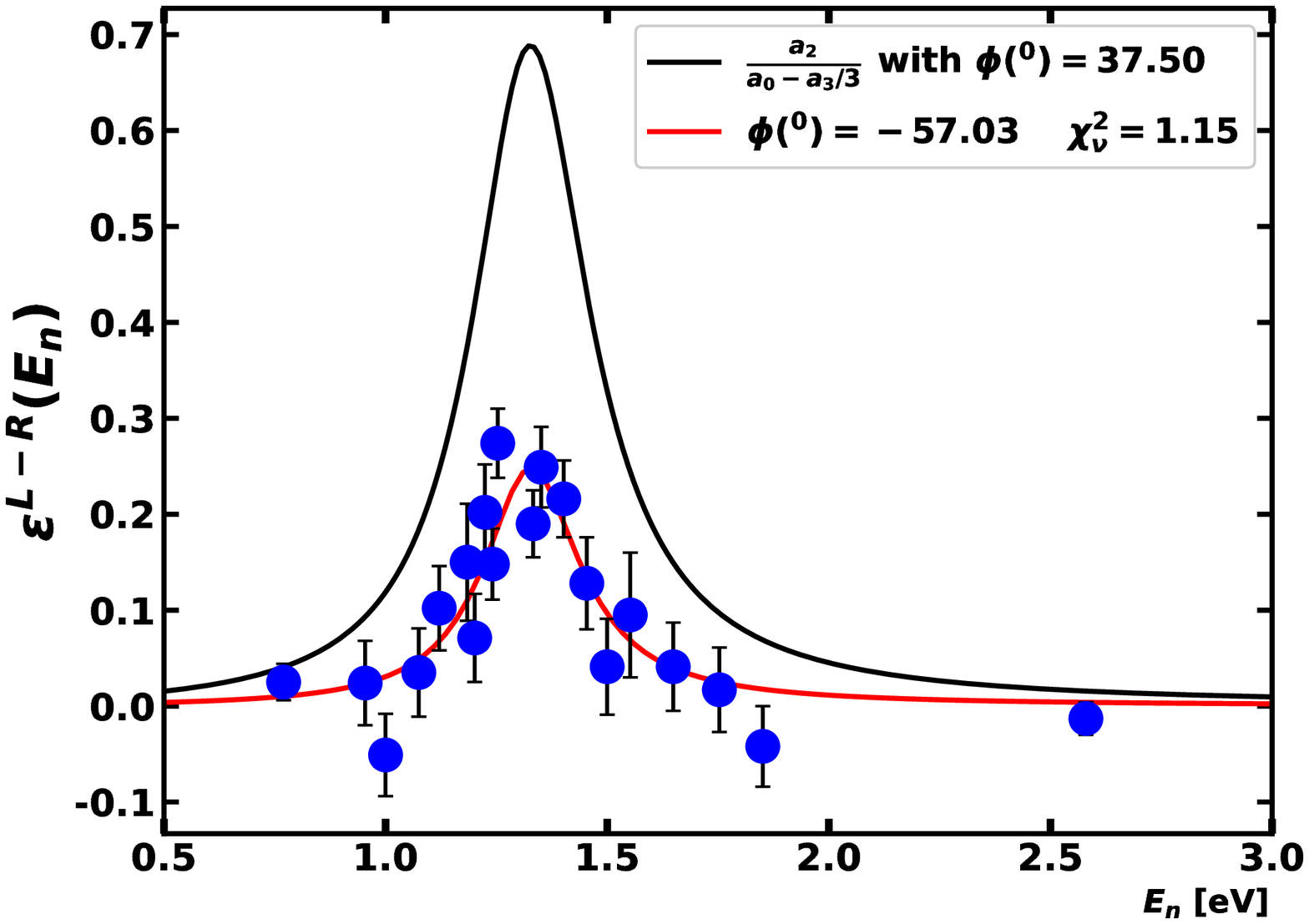}
\caption{Measured values for the left-right asymmetry at the vicinity of the \textit{p}-resonance $E_{p}=1.33$ eV for $^{117}$Sn. Red curves correspond to the best fits: $\phi(^{0})=-57.03$ ($\chi^{2}_{\nu}=1.15$) and $\phi(^{0})=92.88$ ($\chi^{2}_{\nu}=1.15$). Black curves correspond to calculated curves using obtained $\phi$ values from the fitting on measured F-B asymmetry values.}
\label{fig:figure3}
\end{figure}

\begin{figure}[h!]
\includegraphics[width=3.2in]{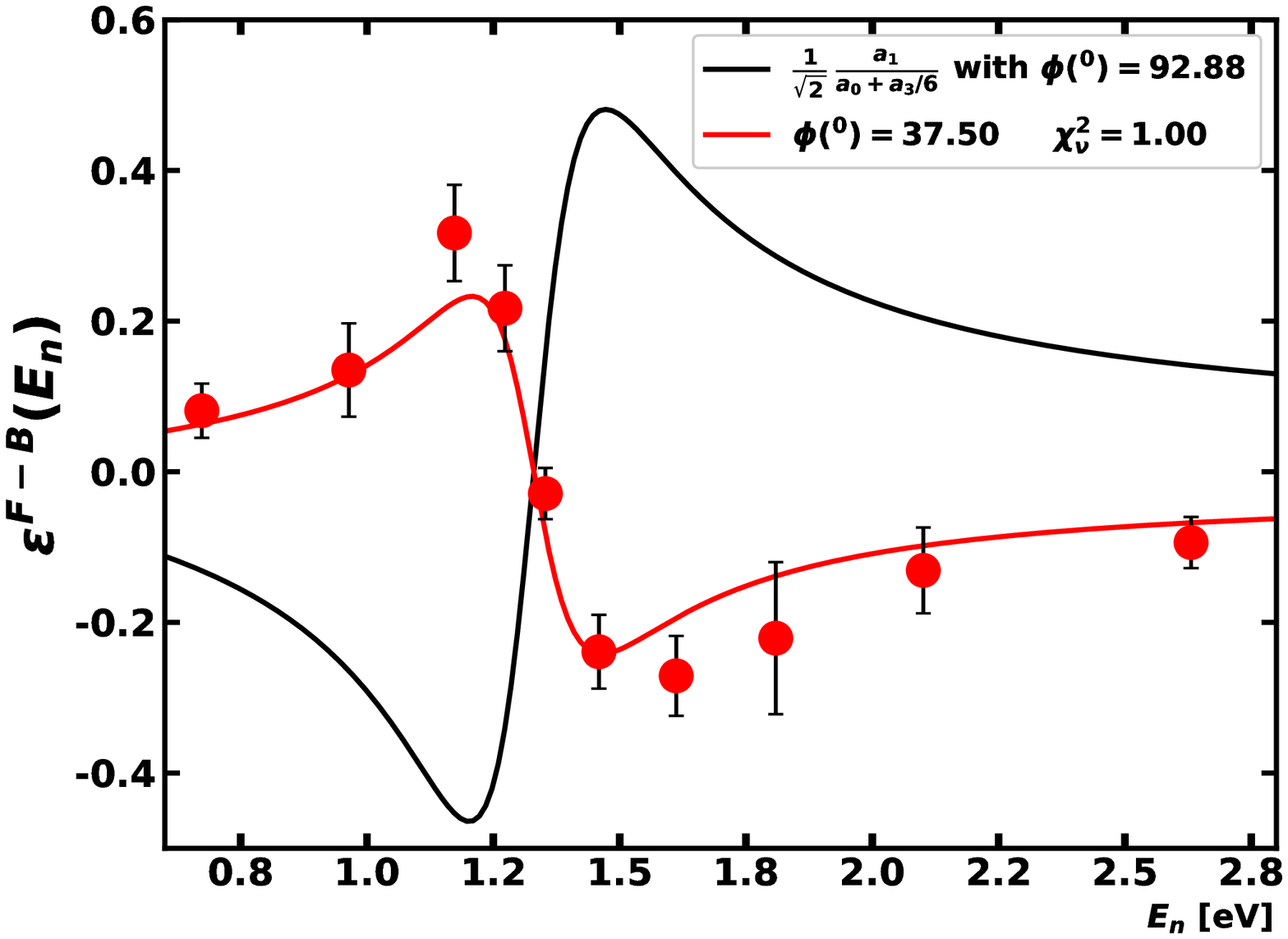}
\includegraphics[width=3.2in]{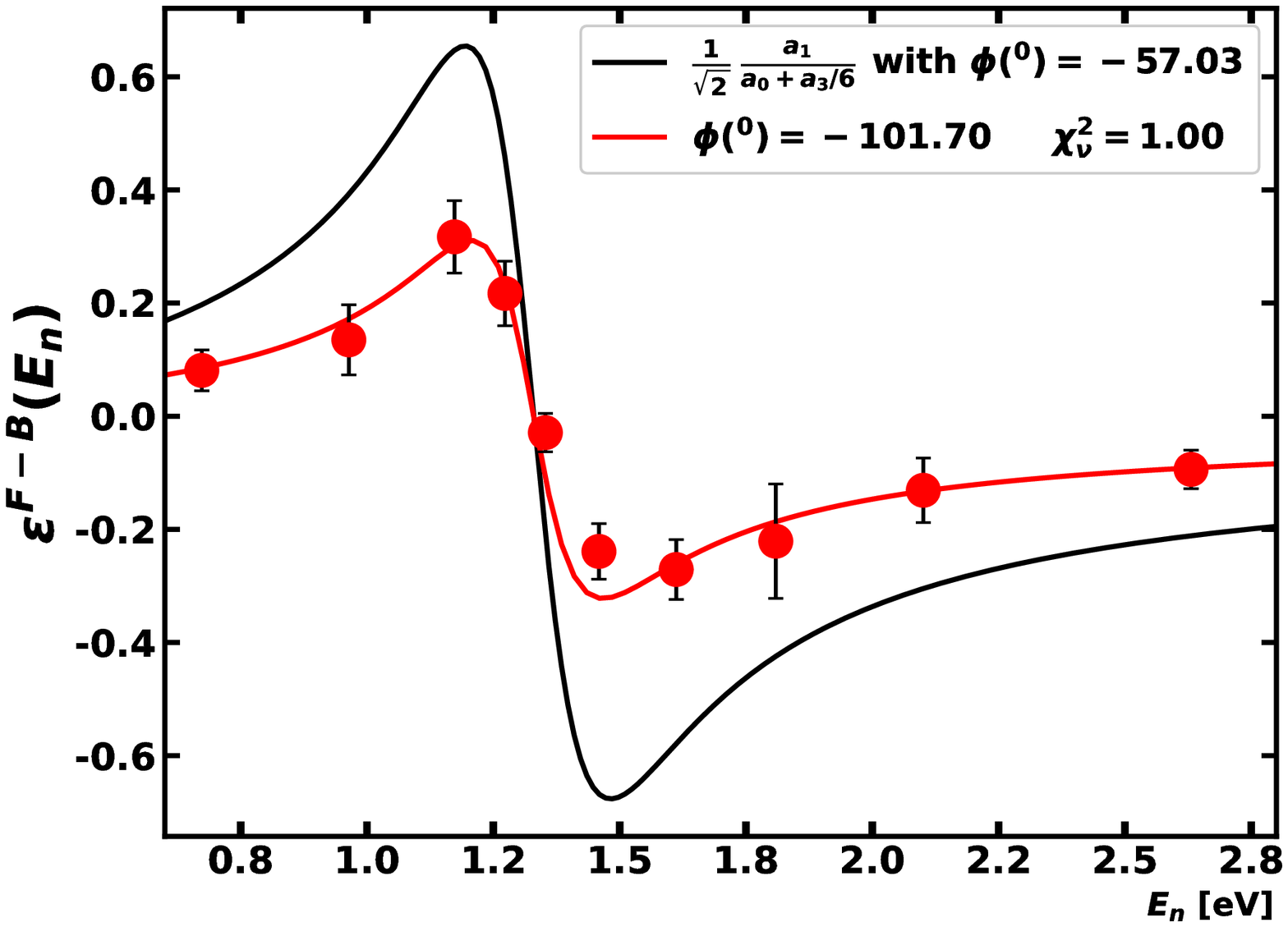}
\caption{Measured values for the forward-backward asymmetry at the vicinity of the \textit{p}-resonance $E_{p}=1.33$ eV for $^{117}$Sn. Red curves correspond to the best fits: $\phi(^{0})=-101.70$ ($\chi^{2}_{\nu}=1.00$) and $\phi(^{0})=37.5$ ($\chi^{2}_{\nu}=1.00$). Black curves correspond to calculated curves using obtained $\phi$ values from the fitting on measured L-R asymmetry values.}
\label{fig:figure4}
\end{figure}

Taking Eqs. (\ref{eq:equation12}) and (\ref{eq:equation13}) spectroscopic parameters from Table (\ref{tab:table1}) in a two-level approximation, we find ourselves in exactly the same situation reported by several authors (refs. \cite{Alfimenkov1985,Barabanov,Sharapov}). We cannot find a unique $\phi$ value that allows us to reproduce both experimental results on $\epsilon^{L-R}$ and $\epsilon^{F-B}$, as can be seen in Figs. (\ref{fig:figure3}) and (\ref{fig:figure4}); for a given $\phi$ value, we obtain either magnitude or sign problems. 

In view of the lack of information on the signs of neutron ($\gamma^{n}_{s(p)}$) and gamma ($\gamma^{\gamma}_{s(p)}$) transition amplitudes for the s(p)-compound state, we propose to assume a negative sign in Eqs. (\ref{eq:equation12}) and (\ref{eq:equation13}) to see if the additional spectroscopic data can be fit. After fitting the experimental values for $\epsilon^{L-R}$ ($\epsilon^{F-B}$) we obtain two possible values for $\phi$ (red curves in figures). With these $\phi$ values we can try to reproduce the measured values for the other asymmetry $\epsilon^{F-B}$ ($\epsilon^{L-R}$) evaluating Eqs. (\ref{eq:equation12}) and (\ref{eq:equation13}) with the opposite sign (black curves in figures).

\begin{figure}[h!]
\includegraphics[width=3.2in]{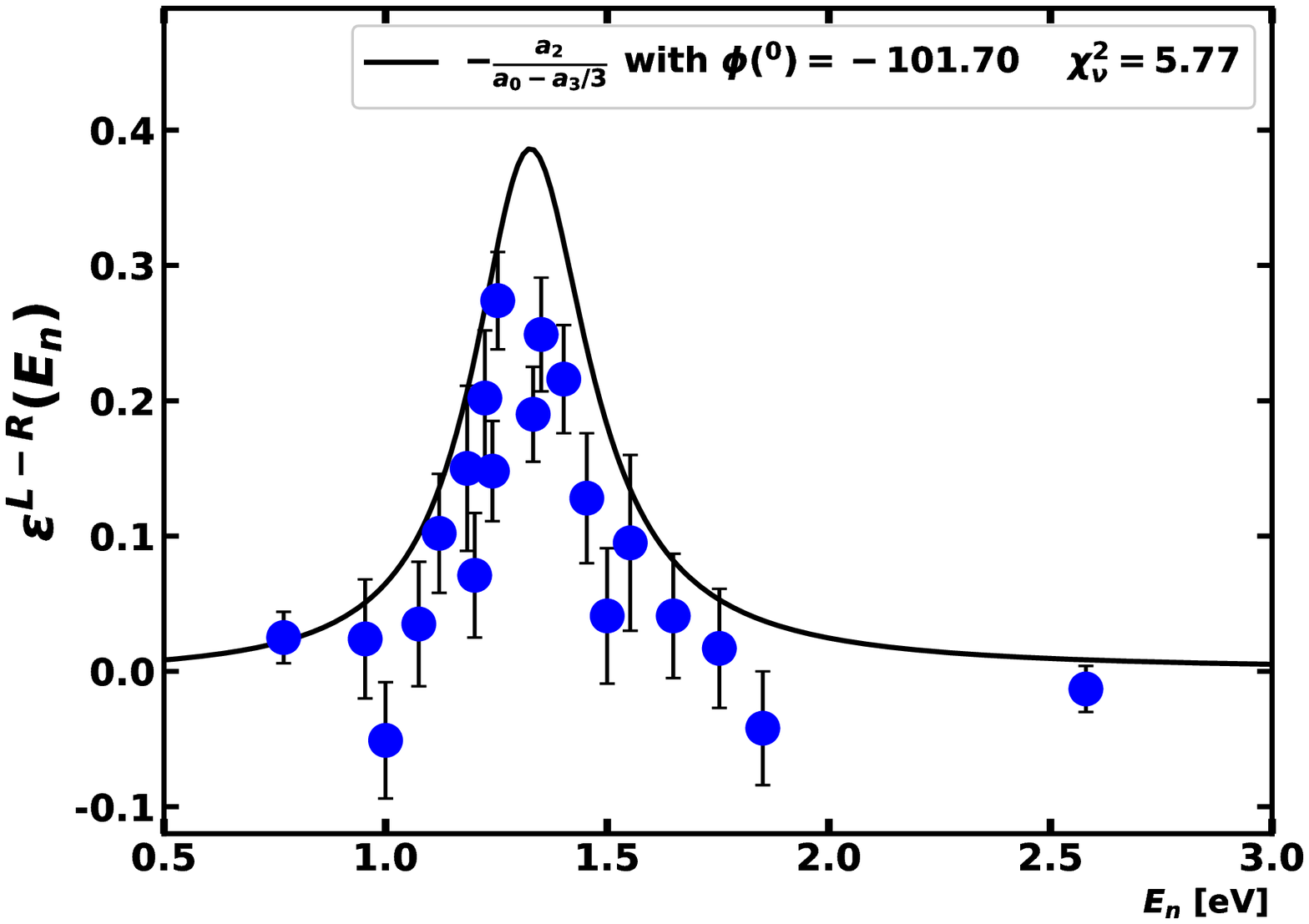}
\includegraphics[width=3.2in]{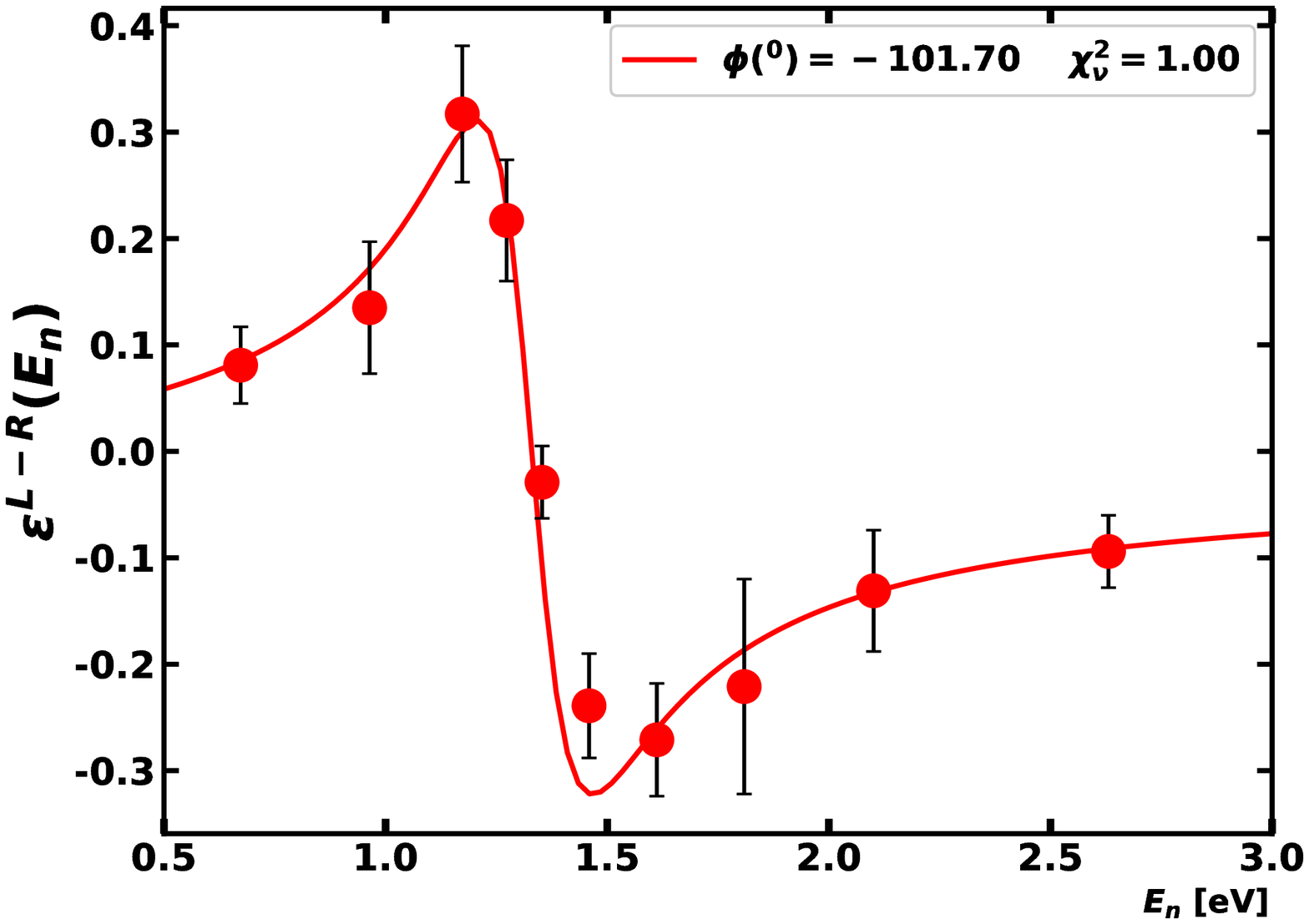}
\caption{Measured values for the left-right and forward-backward  asymmetries at the vicinity of the \textit{p}-resonance $E_{p}=1.33$ eV for $^{117}$Sn. Red curve corresponds to the best fit: $\phi(^{0})=-101.70$ ($\chi^{2}_{\nu}=1.00$). Black curve corresponds to the calculated curve using the fitted $\phi$ value from F-B asymmetry with $\chi^{2}_{\nu}=5.771$.}
\label{fig:figure5}
\end{figure}

Taking 

\begin{equation}
\epsilon^{F-B}(E,\theta=45^{0})=\frac{1}{\sqrt{2}}\frac{a_{1}}{a_{0}+a_{3}/6} \qquad
    \mathrm{and} \qquad
    \epsilon^{L-R}(E)=-\frac{a_{2}}{a_{0}-a_{3}/3},
   \label{eq:equation18}
\end{equation}

we obtain Fig. (\ref{fig:figure5}). 

\newpage

From the figure we are able to reproduce experimental values for these two measured P-even asymmetries with a unique $\phi$ value. On the other hand, taking

\begin{equation}
\epsilon^{L-R}(E)=\frac{a_{2}}{a_{0}-a_{3}/3} \qquad
    \mathrm{and} \qquad
    \epsilon^{F-B}(E,\theta=45^{0})=-\frac{1}{\sqrt{2}}\frac{a_{1}}{a_{0}+a_{3}/6},
\label{eq:equation19}
\end{equation}

we obtain the same result. Notice here that we do not have problems with the sign of the P-even effects or their order of magnitude. In these cases the fitting of a single asymmetry gives us a value for $\phi$ that fits both measured observables.  

\begin{figure}[h!]
\includegraphics[width=3.2in]{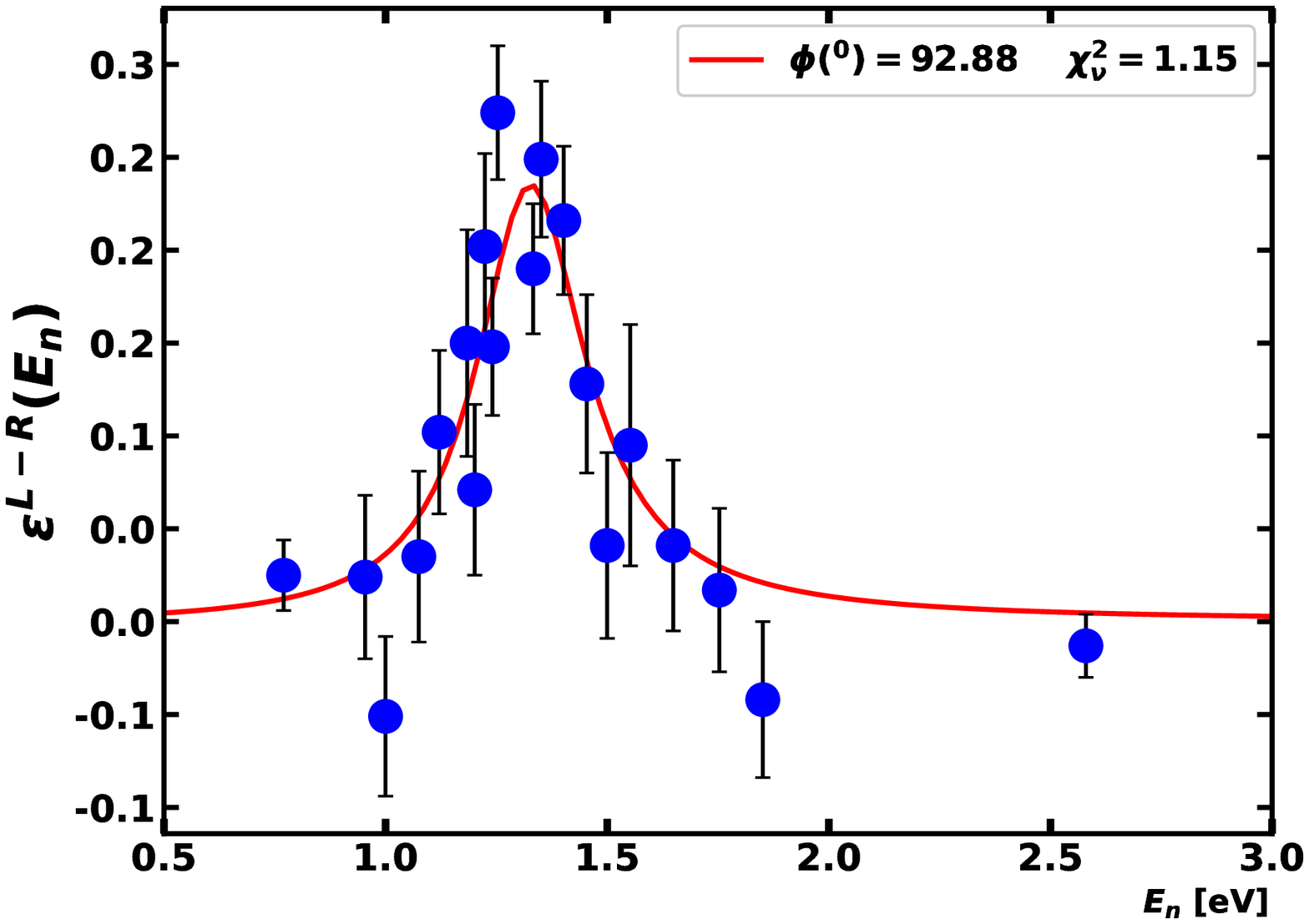}
\includegraphics[width=3.2in]{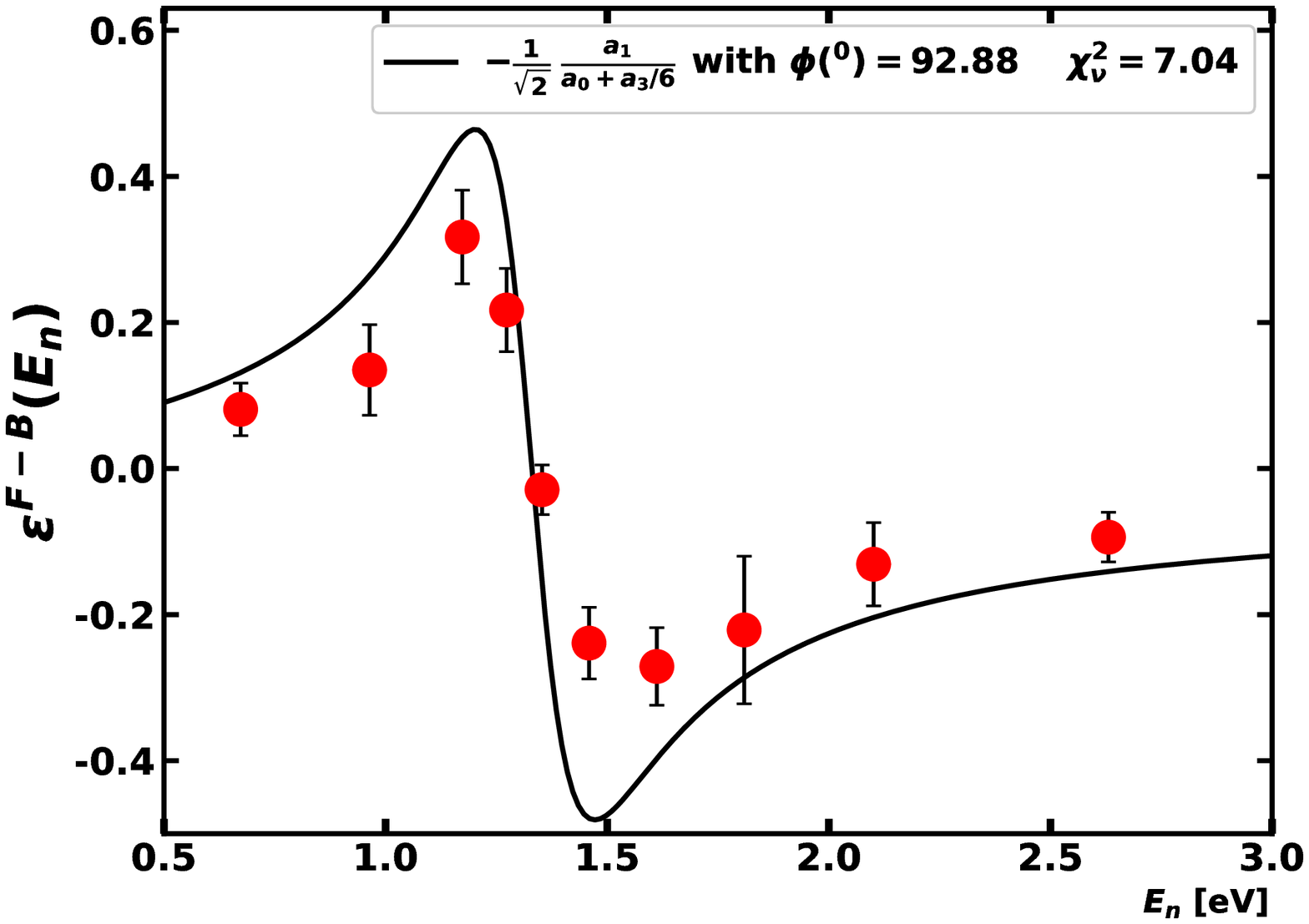}
\caption{Measured values for the left-right and forward-backward  asymmetries at the vicinity of the \textit{p}-resonance $E_{p}=1.33$ eV for $^{117}$Sn. Red curve corresponds to the best fit: $\phi(^{0})=92.88$ ($\chi^{2}_{\nu}=1.15$). The black curve corresponds to the calculated curve using this fitted $\phi$ value from L-R asymmetry with $\chi^{2}_{\nu}=7.04$.}
\label{fig:figure7}
\end{figure}

In the isolated resonance regime it is common to take only one s-resonance but in the more general case we can have more than one s-resonances contributing to the P-even effect at the p-resonance energy. The mixing of the p-state with those s-resonances with the largest $\Gamma^{n0}_{s}/(E_{s})^2$ values (the positive s-resonance at 38.8 eV and the sub-threshold state at -29.2 eV) should make the dominant contribution to the asymmetries~\cite{Barabanov}. Below we investigate whether or not the inclusion of this positive s-resonance in the fits lowers the $\chi^{2}_{\nu}$.

\subsubsection{Three-level approximation}

\begin{figure}[h!]
\includegraphics[width=3.2in]{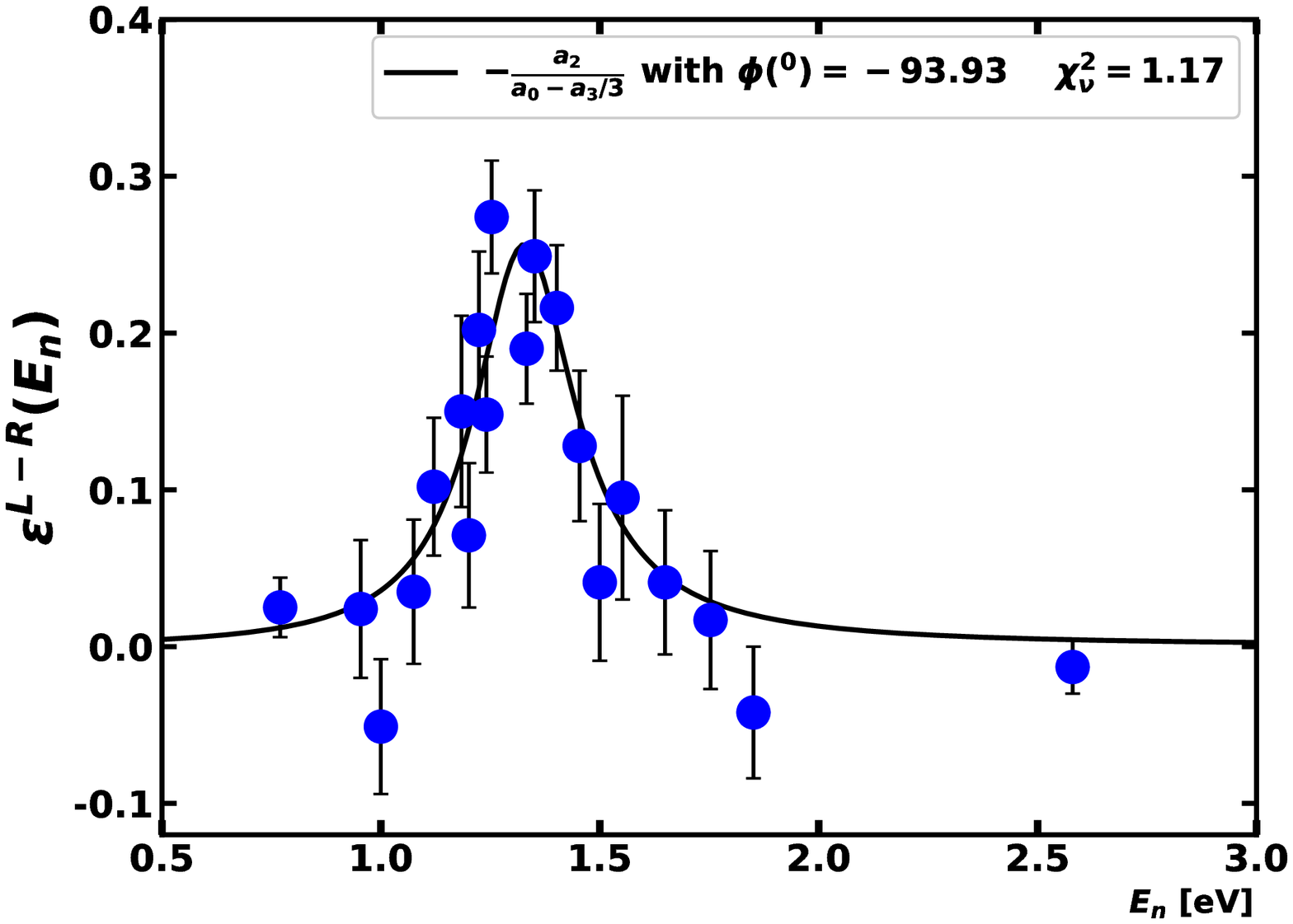}
\includegraphics[width=3.2in]{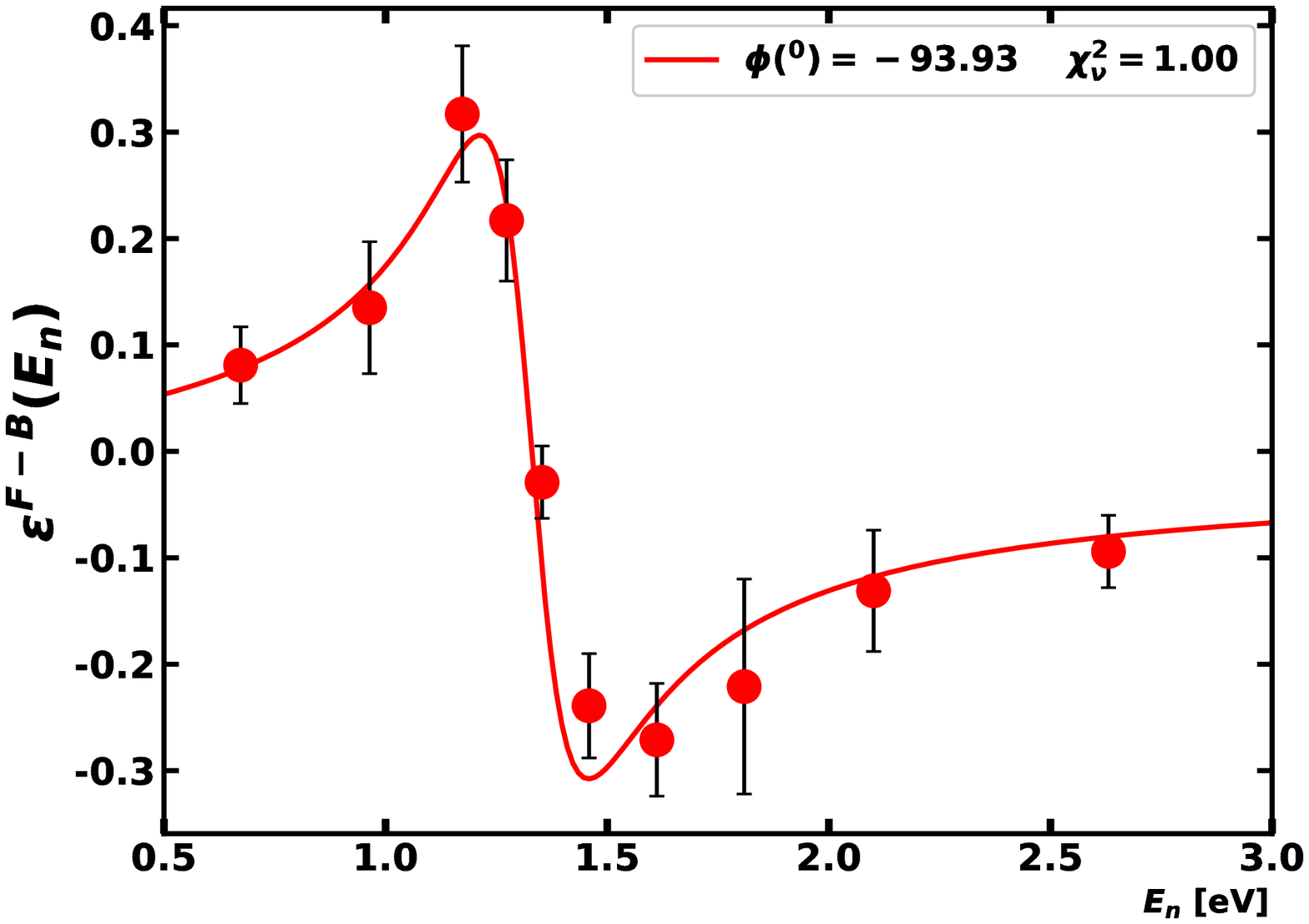}
\caption{Measured values for the left-right and forward-backward  asymmetries at the vicinity of the \textit{p}-resonance $E_{p}=1.33$ eV for $^{117}$Sn. Red curve corresponds to the best fit: $\phi(^{0})=-93.93$ ($\chi^{2}_{\nu}=1.00$). Black curve corresponds to the calculated curve using this fitted $\phi$ value from F-B asymmetry with $\chi^{2}_{\nu}=1.17$.}
\label{fig:figure9}
\end{figure}

\begin{figure}[h!]
\includegraphics[width=3.2in]{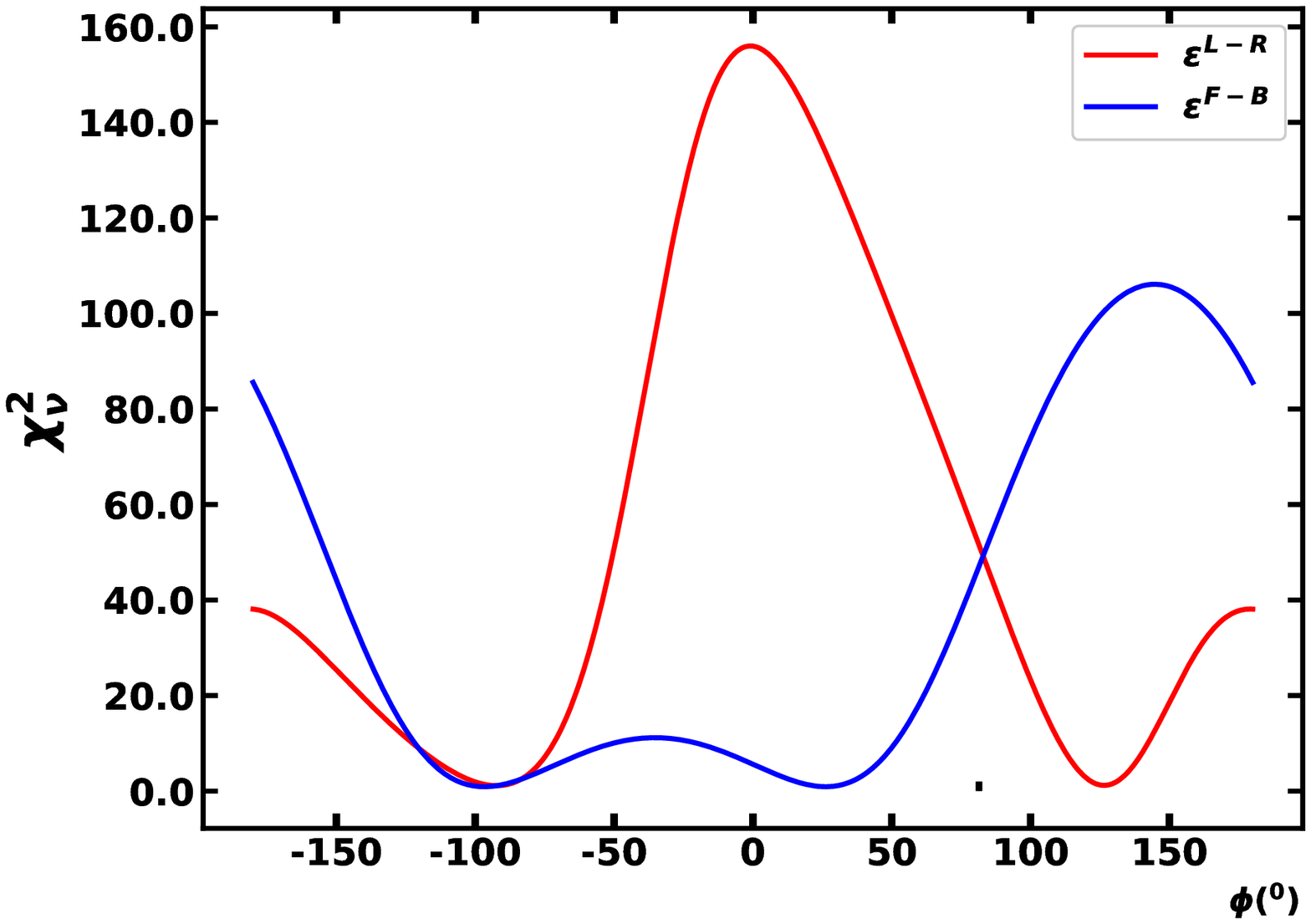}
\includegraphics[width=3.2in]{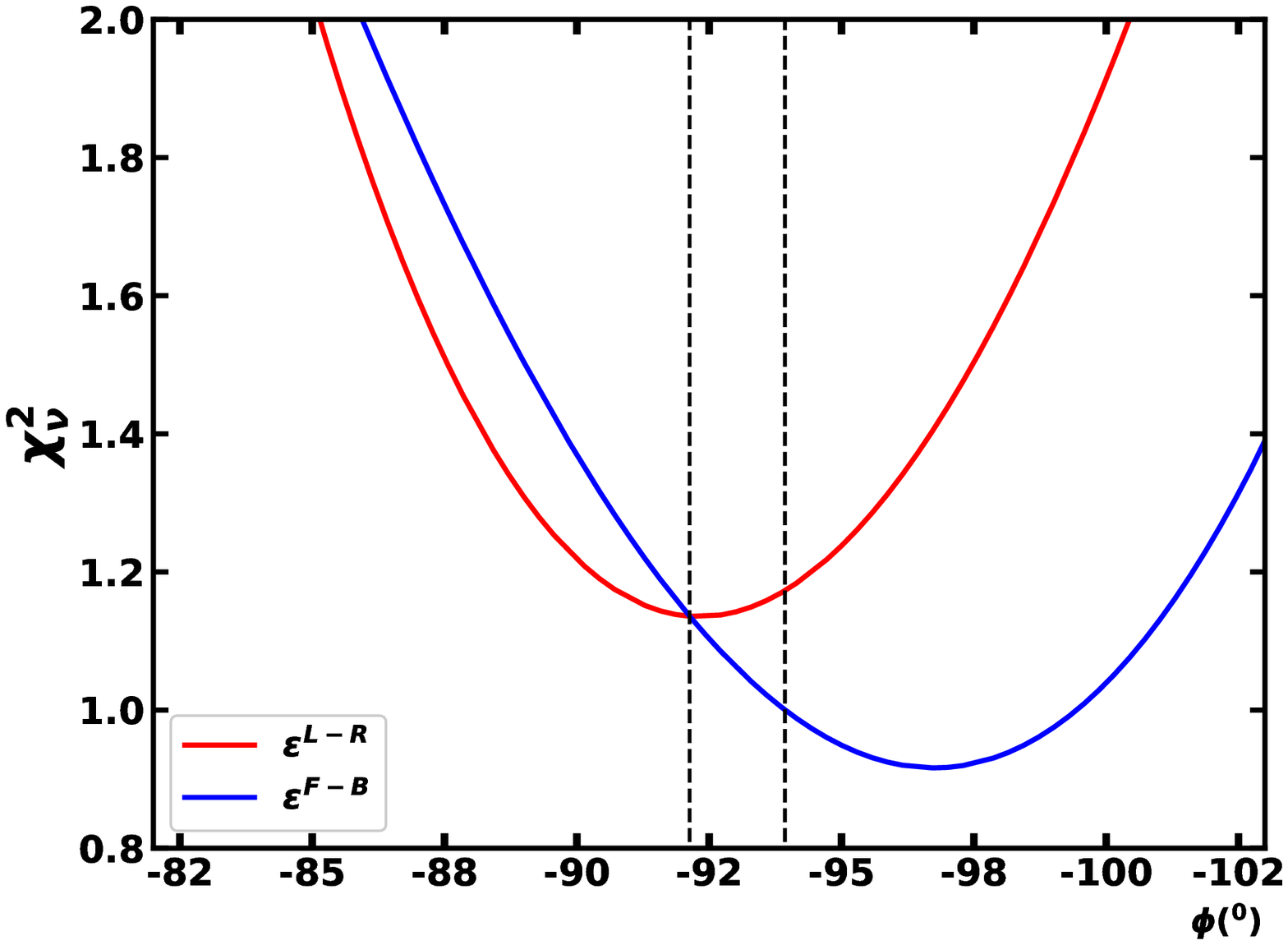}
\caption{$\chi^{2}_{\nu}$ as a function of the parameter $\phi$ (a negative s-resonance at -29.2 eV, a p-resonance at 1.33 eV, a positive s-resonance at 38.8 eV and considering three-level approximation, destructive interference and spectroscopic parameters by Alfimenkov).}
\label{fig:figure10}
\end{figure}

Considering \ref{eq:equation18}, spectroscopic parameters from Table (\ref{tab:table1}), three-level approximation (one p- and two s-resonances) and destructive interference, we obtain Figs. (\ref{fig:figure9}) and (\ref{fig:figure10}). The value of $\phi$ from this fit does not change much from that obtained in the two-level approximation analysis, but the quality of the fit as measured by $\chi^{2}_{\nu}$ improves noticeably. Note carefully that this success occurs in spite of the fact that we did not treat the positive resonance parameters as adjustable fit parameters: there is no change from their accepted values in the scientific literature. The only additional new undetermined ``parameter" involved in this three resonance fit is the relative sign of the positive s-wave resonance term compared to that from the subthreshold resonance. As we pointed out above, such signs are not determined by other measurements and must be treated as free parameters in our case.

From Fig. (\ref{fig:figure10}), for the interval 

\begin{equation}
    \phi \in [-93.93;-92.13],
 \label{eq:equation22}
\end{equation}

we have

\begin{equation}
\begin{split}
    \chi^{2}_{\nu_{LR}}  \in [1.17;1.14], \\
     \chi^{2}_{\nu_{FB}} \in [1.00;1.14].
\end{split}
 \label{eq:equation23}
\end{equation}

\begin{figure}[h!]
\includegraphics[width=3.2in]{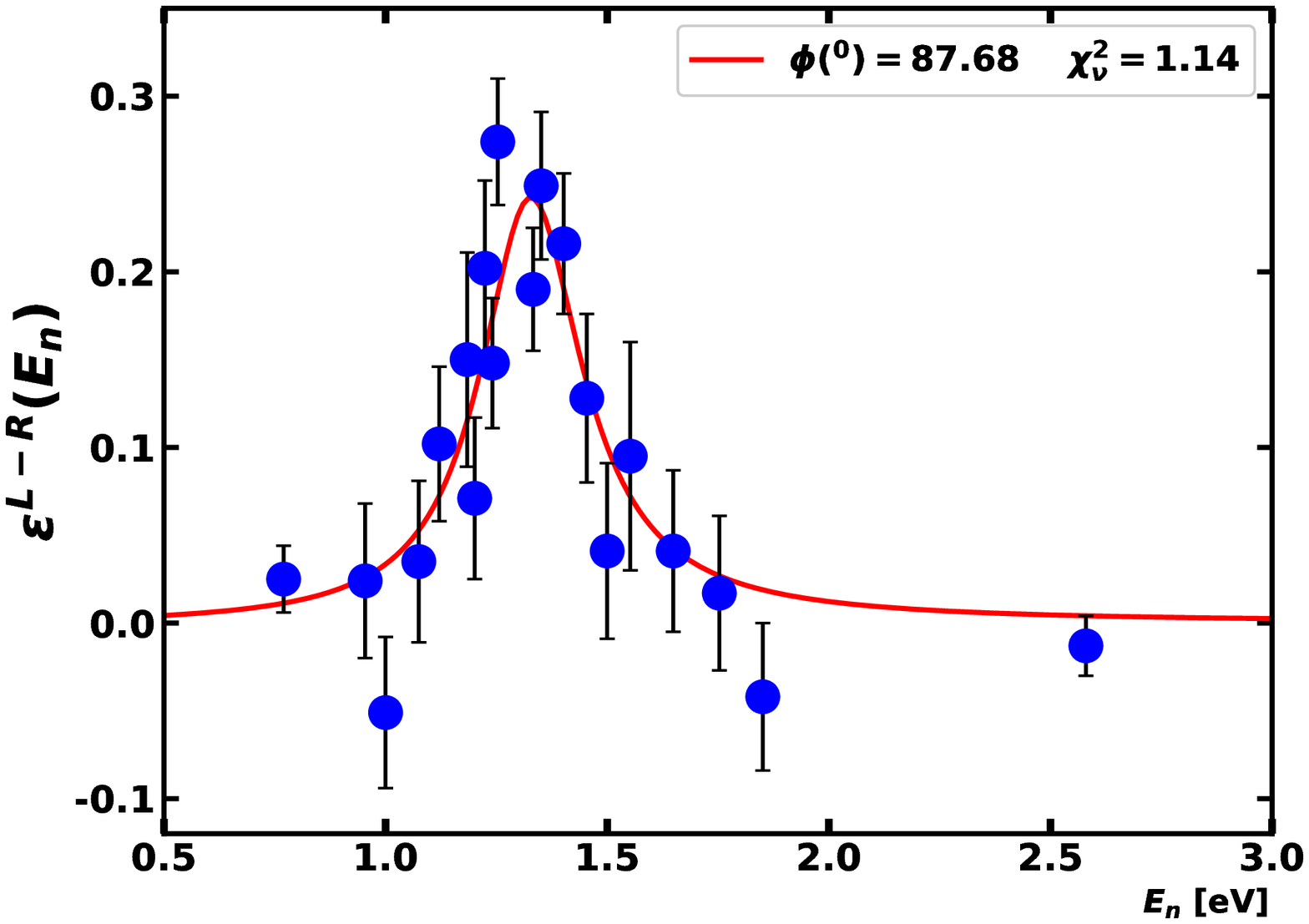}
\includegraphics[width=3.2in]{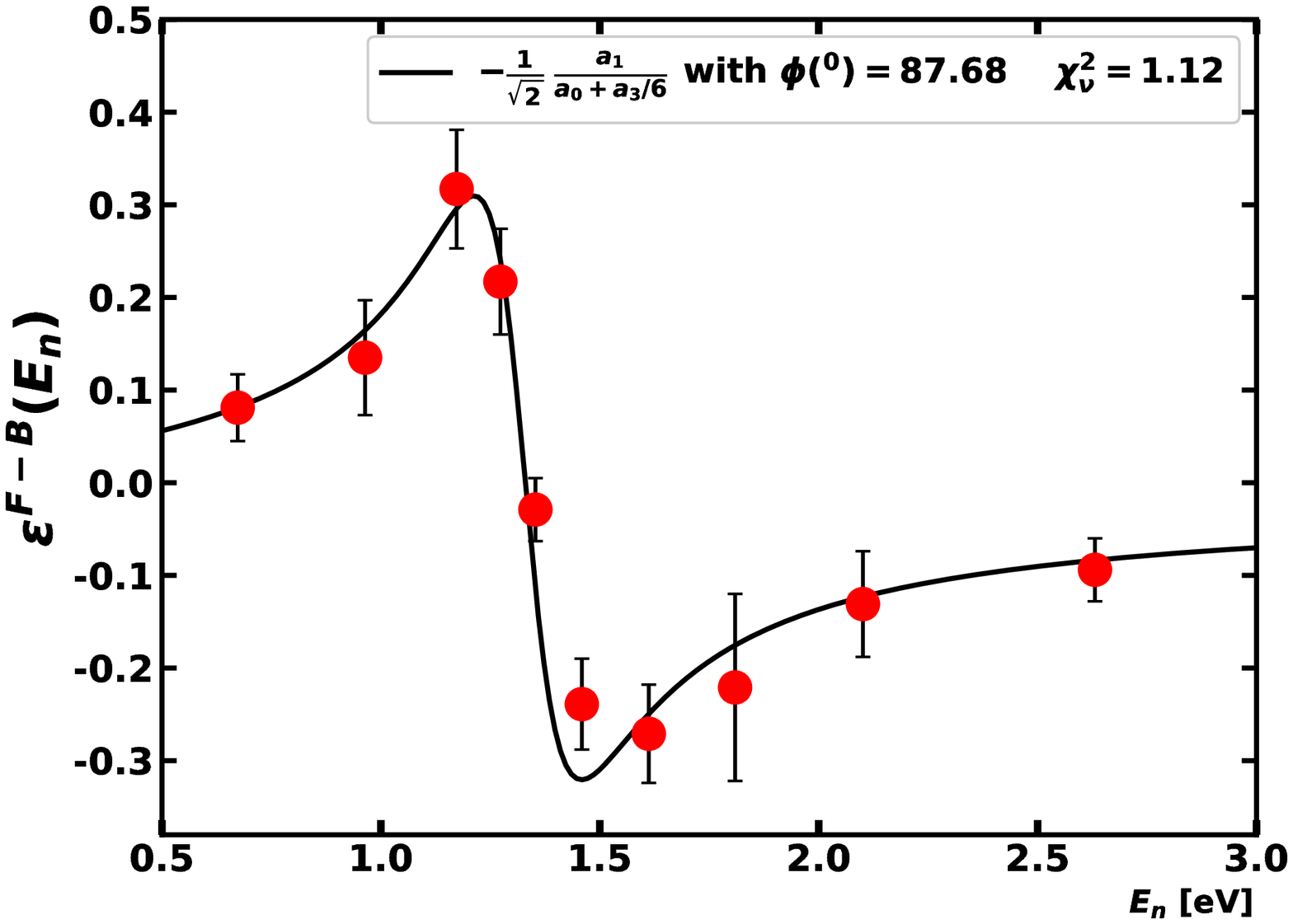}
\caption{Measured values for the left-right and forward-backward  asymmetries at the vicinity of the \textit{p}-resonance $E_{p}=1.33$ eV for $^{117}$Sn. Red curve corresponds to the best fit: $\phi(^{0})=87.68$ ($\chi^{2}_{\nu}=1.14$). Black curve corresponds to the calculated curve using this fitted $\phi$ value from L-R asymmetry with $\chi^{2}_{\nu}=1.12$.}
\label{fig:figure11}
\end{figure}

\begin{figure}[h!]
\includegraphics[width=3.2in]{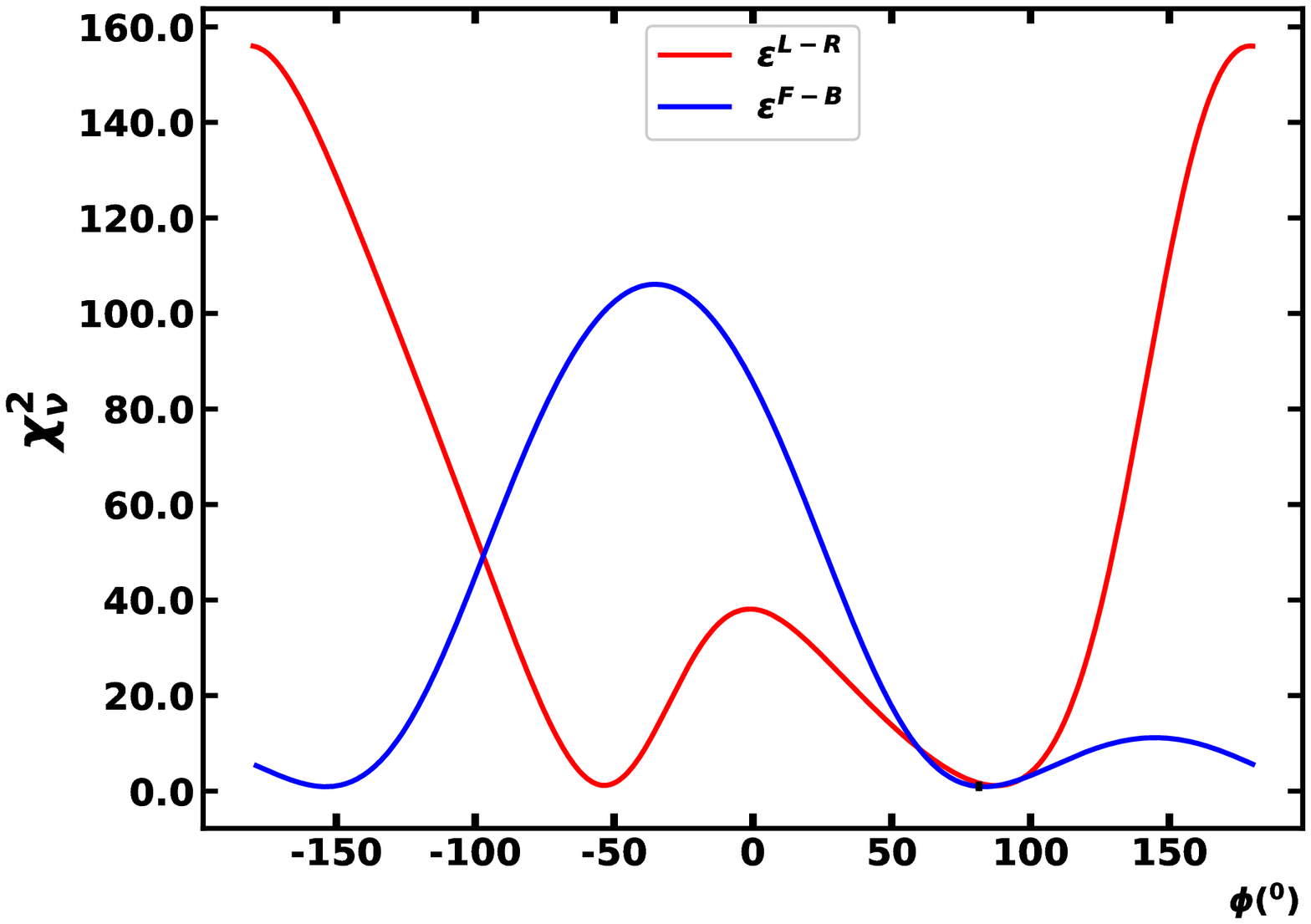}
\includegraphics[width=3.2in]{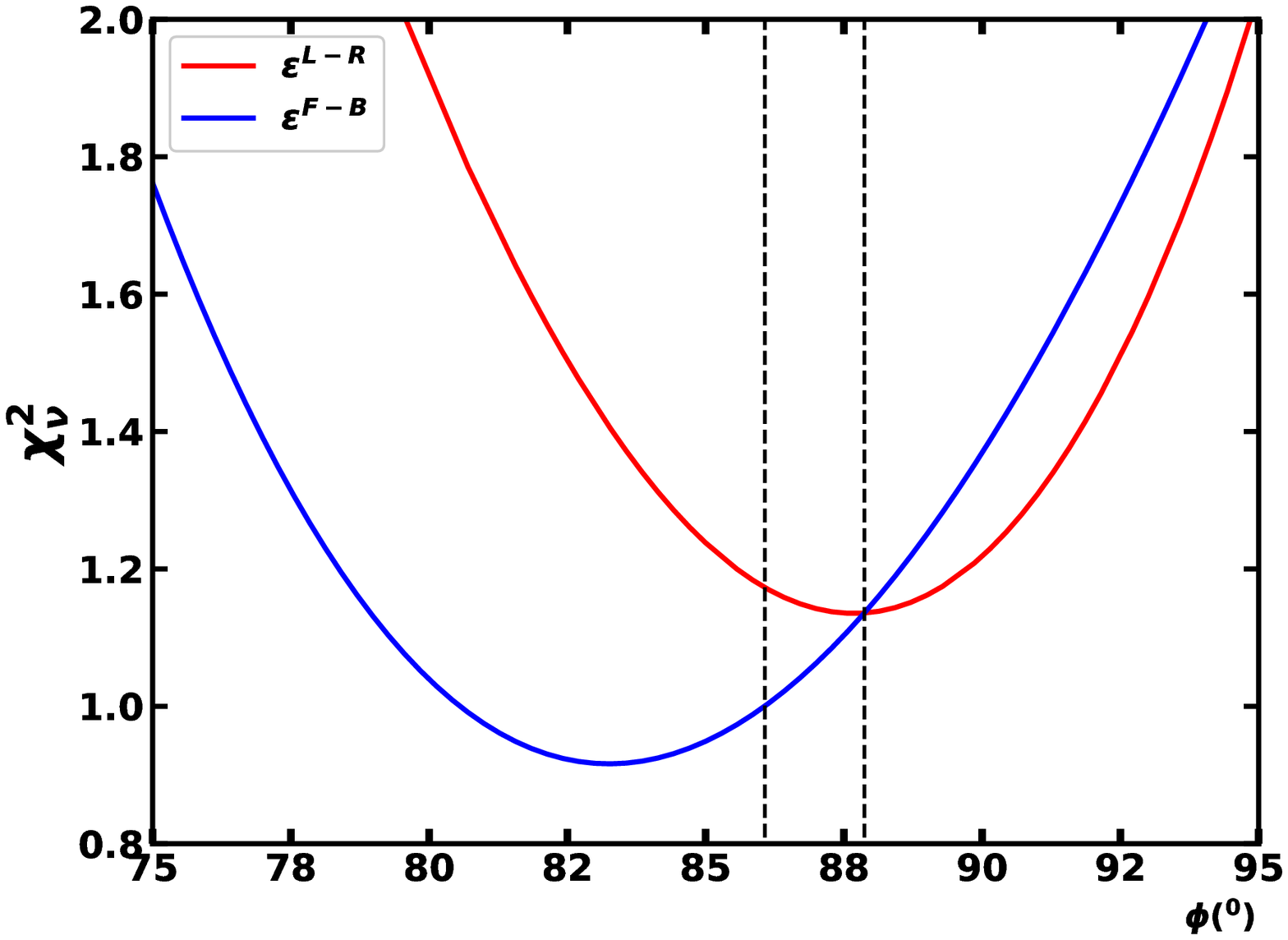}
\caption{$\chi^{2}/ndf$ as a function of the parameter $\phi$ (a negative s-resonance at -29.2 eV, a p-resonance at 1.33 eV, a positive s-resonance at 38.8 eV and considering three-level approximation, destructive interference and spectroscopic parameters by Alfimenkov).}
\label{fig:figure12}
\end{figure}

\begin{figure}[h!]
\includegraphics[width=5.0in]{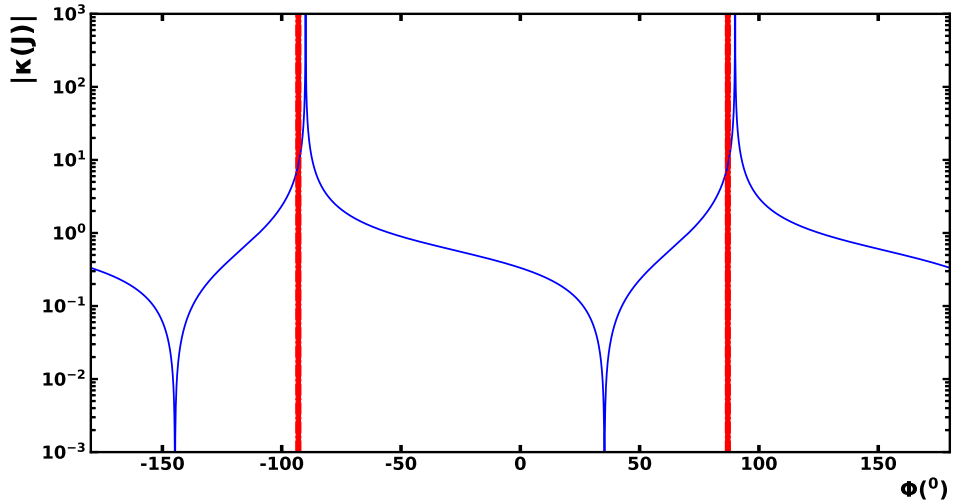}
\caption{Values obtained for the parameter $\kappa(J)$ (Eq. \ref{eq:kappa_J}) in the present analysis. Besides there is no constrain on $\kappa(J)$ values derived from any formalism, values between $10^{-1}$ and $10$ are expected.}
\label{fig:kappa}
\end{figure}

On the other hand, taking Eq. (\ref{eq:equation19}) we obtain

\begin{equation}
    \phi \in [86.07;87.87],
 \label{eq:equation24}
\end{equation}

that also satisfy Eq. (\ref{eq:equation23}). From Figs. (\ref{fig:figure11}) and (\ref{fig:figure12}) we can see an improvement in the evaluated $\chi^{2}_{\nu}$ values for $\epsilon^{L-R}$ and $\epsilon^{F-B}$. This means that the positive s-resonance makes an important contribution to the measured asymmetries at the p-resonance energy. From the former two intervals $\phi\in\{[-93.93;-92.13]\cup[86.07;87.87]\}$ we obtain Fig. (\ref{fig:kappa}) 

\begin{equation*}
\kappa\in[6.53;12.34].\end{equation*}

%%%%%%%%%%%%%%%%%%%%%%%%%%%%%

The measurement of the parameter $\Gamma_{p}$ is not an easy experimental task. Alfimenkov measured the parameter $\Gamma^{\gamma}$ through (n,$\gamma$) reactions at 1.33 eV in $^{117}$Sn. He reported a first value $\Gamma^{\gamma}_{p}=0.23\pm0.02$ eV~\cite{Alfimenkov1983}. Since $^{117}$Sn is a non-fissile nucleus and the total cross section is practically equal to the radiative capture cross section at these very low energies ($\Gamma^{n} \ll \Gamma^{\gamma}$) we can make the approximation $\Gamma_{p}\approx \Gamma^{\gamma}_{p}$. Later Smith \textit{et al.}~\cite{Smith1999} reported a lower value $\Gamma^{\gamma}_{p}=0.148 \pm 0.010$ eV which then was adopted into the next edition of Mughabghab~\cite{Mughabghab2006}. Usually $\Gamma_{p}=\Gamma_{s}\approx0.1$ eV \cite{Bunakov1981} for most resonances. We can also find a lower value $\Gamma^{\gamma}_{p}=0.180\pm 0.018$ eV by Alfimenkov in references \cite{1,3,Alfimenkov1985-JINR}. We fit the data using all of these values for $\Gamma^{\gamma}_{p}$. We obtain the best fits with $\Gamma^{\gamma}_{p}=0.148$ eV. There are additional good reasons to believe that this narrower width is likely to be closer to the correct value based on the differences of the source and apparatus properties used in these different measurements. The beam and measurement apparatus used in~\cite{Smith1999} were a qualitative improvement over that available to~\cite{Alfimenkov1983}. The intrinsic spread of the neutron energies from the water moderator at the spallation source used in~\cite{Smith1999} is narrower than that from the neutron source used in~\cite{Alfimenkov1983}, and the time resolution of both the $^{10}$B-loaded liquid scintillation neutron detector used in transmission and the pure CsI gamma detector array used in (n, $\gamma$) mode in~\cite{Smith1999} were sharper than the detectors available to~\cite{Alfimenkov1983}. We suspect that these instrumental improvements could be the reason why Smith \textit{et al.} reported a narrower resonance width $\Gamma^{\gamma}_{p}$. Combined with the high statistical accuracy obtained in this work, the Smith \textit{et al.} result domiantes the reported errors on the resonance energy and width.

The recent measurements of 
 $\gamma$-rays angle distribution with unpolarized neutrons  at 1.33 eV in $\mathrm{n}+^{117}$Sn \cite{Koga2022} are consistent with the values 
 $\phi_{1} \approx -88^{0}$ and $\phi_{2} \approx 18^{0}$. 
 While the measurements of the left-right  asymmetry \cite{Endo,endo_2023} reported $\phi_{1}\approx -2.0^{0}$ and $\phi_{2}\approx 40.9^{0}$, with the value of the left-asymmetry $A_{LR}=1.07\pm0.23$.
 These results do not agree with each others, and with the obtained interval in the present analysis $\phi\in\{[-93.93;-92.13]\cup[86.07;87.87]\}$ that guarantees $\chi^{2}_{\nu}\in[1.00;1.17]$. 
 This indicates that more precise experiments are required to resolve the possible source of inconsistency, which is more likely related to low statistics in the previous measurements.

\subsection{P-component angular anisotropy $\epsilon^{a}_{p}$ and $t^{2}(\theta,E_{p})$}

\begin{figure}[h!]
\includegraphics[width=4.5in]{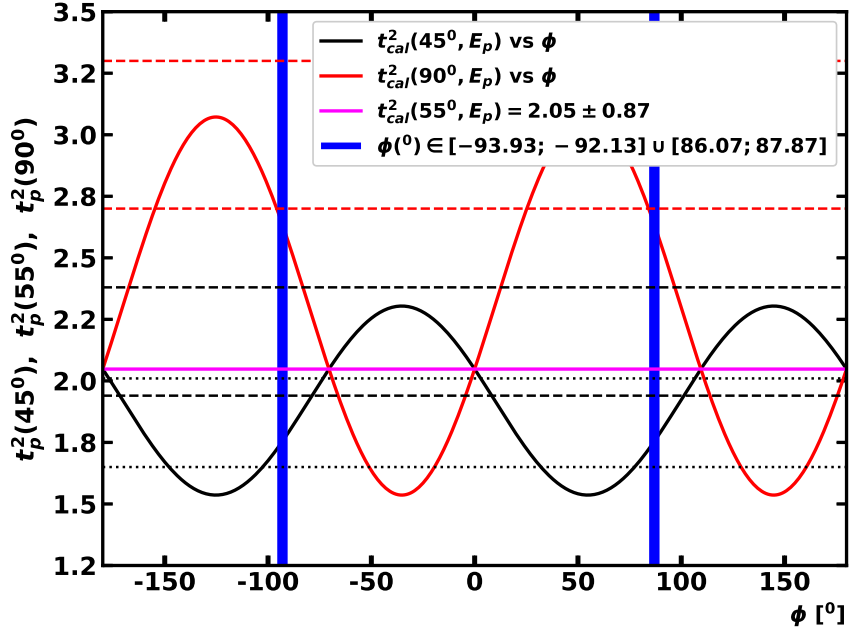}
\caption{Calculated values for spectroscopic parameters $t^{2}_{45^{0}}(E_{p})$, $t^{2}_{55^{0}}(E_{p})$  and $t^{2}_{90^{0}}(E_{p})$ using three-level approximation, destructive interference, Flambaum-Sushkov formalism,  and values by Alfimenkov. Experimental values $t^{2}_{exp}(90^{0},E_{p})=3.0\pm 0.3$ (red dashed line), $t^{2}_{exp}(45^{0},E_{p})=2.16\pm 0.22$ (black dashed line) and $t^{2}_{exp}(45^{0},E_{p})=1.83\pm 0.18$ (black dotted line) are also shown. There is agreement between the calculated and experimental values if we take into account the values for $\phi$  from $\epsilon^{L-R}$ and $\epsilon^{F-B}$.}
\label{fig:figure13}
\end{figure}

\begin{figure}[h!]
\includegraphics[width=4.5in]{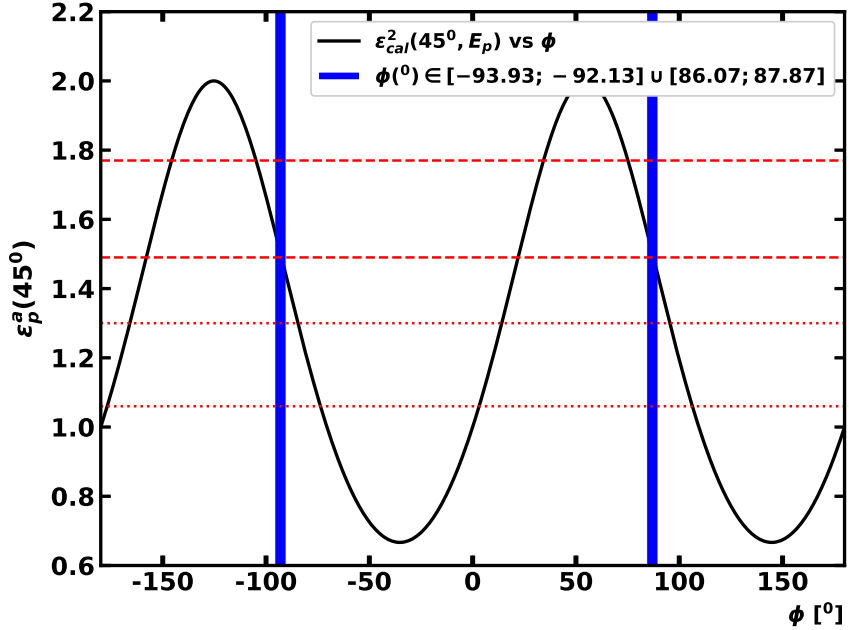}\caption{Experimental and theoretical values for the spectroscopic parameter $\epsilon^{a}_{p}(45^{0})$ using three-level approximation, destructive interference, Flambaum-Sushkov formalism and values by Alfimenkov. Experimental values $\epsilon^{2}_{exp}(45^{0},E_{p})=1.63\pm 0.14$ (red dashed line) and $\epsilon^{2}_{exp}(45^{0},E_{p})=1.18\pm 0.12$ (red dotted line) are also shown. There is agreement between the calculated and experimental values if we take into account the values for $\phi$  from $\epsilon^{L-R}$ and $\epsilon^{F-B}$.}
\label{fig:figure14}
\end{figure}

In Table (\ref{tab:table1}) values for $\epsilon^{a}_{p}$ were obtained using Eq. (\ref{eq:equation17}) and measured values for $t^{2}_{p}(\theta)$. Taking into account directly experimental values for $t^{2}_{p}(45^{0})$, $t^{2}_{p}(55^{0})$, $t^{2}_{p}(90^{0})$ and evaluated ones for $t^{2}_{cal}(\theta,E_{p})$ from spectroscopic parameters from Table (\ref{tab:table1}) we obtain Figs. (\ref{fig:figure13}) and (\ref{fig:figure14}).

From these figures  we can see that $\phi$ is consistently determined from $\epsilon^{L-R}_{p}$ and $\epsilon^{F-B}_{p}$ data when analyzed in the three-level approximation (destructive interference) and Flambaum-Sushkov formalism. In this case, we have agreement considering only the value $\epsilon^{a}_{p}=1.63 \pm 0.14$. The fact that we obtain the best result in three-level approximation means that the positive s-resonance at 38.8 eV has an important contribution to the asymmetries at the p-resonance energy $E_{p}=1.33$ eV in $^{117}$Sn.  

It would be no surprise if the addition of more fitting parameters to a theoretical model lowers the chi-squared value of the fit to the data. We emphasize again that in this study we always consider only one fitting parameter $\phi$ both in two- and three-level approximation.

Above we presented our arguments for why we chose to base our main analysis of the data using the subthreshold resonance parameters from the measurements of Alfimenkov, whose values come directly from data in the low-energy neutron range of interest. Nevertheless, we also analyzed the data with values from two different global data evaluations (1981 and 2006 from Mughabghab) which as noted below possess very different values for the subthreshold resonance. We were not able to obtain consistency between experimental and evaluated values for $t^{2}(E_{p},45^{0})$, $t^{2}(E_{p},55^{0})$ and $t^{2}(E_{p},90^{0})$ in two-level approximation from either data evaluation, nor was it possible to succeed with spectroscopic values from the 1981 Mughaghab series in three-level approximation. But we could get an internally consistent analysis from the 2006 Mughaghab series in three-level approximation (destructive interference). In this case, we obtain

\begin{equation}
    \phi = -96.14 \quad \mathrm{and}  \quad 83.86 ,
\end{equation}

and

\begin{equation}
\begin{split}
     \chi^{2}_{\nu_{LR}}  = 1.59, \\
     \chi^{2}_{\nu_{FB}}  = 1.00.
\end{split}
\end{equation}

These values are not far away from those reported in eqs. (\ref{eq:equation22}) and (\ref{eq:equation24}). In this article we only presented our best results (taking values by Alfimenkov). Alfimenkov's values come directly from his own data in the low-energy neutron range of interest. Clearly more experiments would be very valuable to better constrain the spectroscopic parameters in this system, especially for the resonant sub-threshold state in $^{118}$Sn whose properties are clearly important to describe the data. 
 
%%%%%%%%%%%%%%%%%%%%%%%%%%%%%%%%%%%%%%%%%%%%%%%%%%%%%%%%%%%%%%%%%%%%%%

\section{P-odd effects for $^{117}$Sn}

We now turn to the measurements of P-odd effects near the 1.33 eV p-wave resonance of $^{117}$Sn to see if the existing data is internally consistent with the resonance parameters that describe the P-even data. The expression for the P-odd reaction amplitude in neutron transmission experiments can be written as \cite{Bunakov:1982is}:

\begin{equation}
T_{1}=\frac{e^{i\delta_{p}}\gamma^{n}_{p}}{\sqrt{2\pi}} \frac{1}{(E-E_{p})+i\Gamma_{p}/2} \bra{\Phi_{p}}V_{W}\ket{\Phi_{s}} \frac{1}{(E-E_{s})+i\Gamma_{s}/2}    \frac{e^{i\delta_{s}}\gamma^{n}_{s}}{\sqrt{2\pi}}.    
\end{equation}

This term represents the mechanism that contributes the most to P-odd effects in our energy region of interest.  Under the influence of the P-odd weak interaction ($V_{W}$) a mixing transition  occurs between resonance compound states with opposite parity of the compound nucleus. This state then decays under the influence of the strong interaction. The s-wave neutron capture event also contributes to the reaction~\cite{BUNAKOV1997337}.

The P-odd amplitude expressions for (n,$\gamma$) reactions are given in \cite{Flambaum:1983ek}:

\begin{equation*}
f_{3}=-\frac{1}{2k}\sum_{sp}\frac{\bra{f,\gamma}H_{EM}\ket{p}\bra{p}H_{W}\ket{s}\bra{s}H_{s}\ket{n}}{\left(E-E_{p}+i\Gamma_{p}/2\right)\left(E-E_{s}+i\Gamma_{s}/2\right)},
\end{equation*}

\vspace{0.5cm}

\begin{equation*}
f_{4}=-\frac{1}{2k}\sum_{sp}\frac{\bra{f,\gamma}H_{EM}\ket{s}\bra{s}H_{W}\ket{p}\bra{p}H_{s}\ket{n}}{\left(E-E_{s}+i\Gamma_{s}/2\right)\left(E-E_{p}+i\Gamma_{p}/2\right)}.
\end{equation*}

From these reaction amplitudes we obtain in two-level approximation at the thermal energy

\begin{equation}
\frac{P(E_{p})}{P(E_{th})}=4\frac{\sigma_{tot}(E_{th})}{\sigma_{tot}(E_{p})}\frac{(E_{p}-E_{s})\Gamma_{p}\left[(E_{th}-E_{s})^2+\Gamma^{2}_{s}/4\right]\left[(E_{th}-E_{p})^2+\Gamma^{2}_{p}/4\right]}{\left[(E_{th}-E_{s})\Gamma_{p}+(E_{th}-E_{p})\Gamma_{s}\right]\left[(E_{p}-E_{s})^2+\Gamma^{2}_{s}/4\right]\Gamma^{2}_{p}}.
\label{eq:PpPth}
\end{equation}

At the p-resonance energy we have

\begin{equation}
\begin{split}
\frac{P(E_{p})}{d\phi/dz(E_{th})}=\frac{2}{N\sigma_{tot}(E_{p})}\frac{(E_{p}-E_{s})\Gamma_{p}\left[(E_{th}-E_{s})^{2}+\Gamma^{2}_{s}/4\right]\left[(E_{th}-E_{p})^{2}+\Gamma^{2}_{p}/4\right]}{[(E_{th}-E_{s})(E_{th}-E_{p})-\Gamma_{s}\Gamma_{p}/4]\left[(E_{p}-E_{s})^{2}+\Gamma^{2}_{s}/4\right]\Gamma^{2}_{p}}.
\end{split}
\end{equation}

In this case
we eliminate the spin factors dependence (because both $P$ and $d\phi/dz$ share the same spin factor) and the weak-mixing
represented by $v$ which means that we should be able to compare well with the P-even results.

To obtain the former two equations we need to take into account the contribution from the nearest s- and p-resonances and the scattering potential in the total cross section expression  \cite{BUNAKOV1983}

\begin{equation}
    \sigma_{tot}(E)=\frac{\pi}{k^{2}}\frac{\Gamma^{n}_{s}\Gamma_{s}}{(E-E_{s})^2+\Gamma^{2}_{s}/4}+\frac{\pi}{k^{2}}\frac{\Gamma^{n}_{p}\Gamma_{p}}{(E-E_{p})^2+\Gamma^{2}_{p}/4}+(kR)^2.
\label{eq:sigmatotal}
\end{equation}

Taking the spectrometric parameters reported in  Table (\ref{tab:table1}), $\sigma(E_{th})=4$ b, $\sigma(E_{p})=1.8$ b, $N=3.7\cdot10^{22}$ cm$^{-3}$, $d\phi/dz(E_{th})=(-36.7\pm2.7)\cdot10^{-6}$ rad/cm and the average $P(E_{th})=(6.7\pm0.5)\cdot10^{-6}$ from reference \cite{1} and evaluating these equations we obtain 

\begin{equation*}
    P(E_{p}) = (4.7
 \pm 0.4
)\times10^{-3} \quad \mathrm{and} \quad P(E_{p}) = (9.2
 \pm 0.7)\times10^{-3},
\end{equation*}

respectively. These calculated values agree with the experimental values reported in  Table (\ref{tab:P-odd_Ref}).

To estimate the weak matrix element $v$ for heavy and medium-heavy nuclei we can use the phenomenological equation \cite{BUNAKOV1983}

\begin{equation}
    v\sim  10^{-4}\sqrt{\bar{D} \,\mathrm{[eV]}},
\end{equation}

where $\bar{D}$ is the average distance between compound resonances. In the following calculations we use $\bar{D}=48\pm6$ eV \cite{Mughabghab1981} to evaluate some P-odd asymmetry for $^{117}$Sn.

Taking spectroscopic parameters from Table (\ref{tab:table1}) we calculate values for these P-odd effects shown in Table (\ref{tab:Calculated-P-odd}). In order to compare absolute values we need to consider the exact expressions for spin factors.

\begin{table}{c c c c }
\caption{Calculated values for P-odd effects ($P$, $d\phi/dz$ and $\alpha_{n,\gamma_{0}}$) at the thermal and $p$-resonance energy for $^{117}$Sn.} 
\begin{tabular}{ccccc}
\hline
\hline
 $P_{p}$ ($10^{-3}$) & $P_{th}$ ($10^{-6}$) & $\alpha_{n,\gamma_{0_{th}}}$ ($10^{-4}$) & $d\phi_{th}/dz$ ($10^{-6}$) (rad/cm) \\
 \hline
  $-(3.0 \pm 0.13)$ & $-(10 \pm 0.69)$ & $-(7.6 \pm 0.92)$ & $-(48.4 \pm 1.94)$\\
  \hline
  \hline
  \label{tab:Calculated-P-odd}
\end{tabular}
  \label{tab:Calc}
\end{table}

In Table (\ref{tab:P-odd_Ref}) we can see measured values for these asymmetries reported in several references. On the sign of these measured asymmetries we observe that it is very common to find discrepancies between references that cite the same experimental reported value. Our reported values in Table (\ref{tab:Calc}) agree pretty well with those shown in Table (\ref{tab:P-odd_Ref}). 

\begin{table}
\caption{Measured P-odd asymmetries at resonant and thermal energies.}
 \begin{tabular}{c c c c c}
 \hline
 \hline
 $\alpha_{n,\gamma_{th}} (10^{-4})$  &     &  $P_{th}$ ($10^{-6}$) &  & \\
 \hline
  8.1 $\pm$ 1.3   & \cite{BUNAKOV1997}  & -(6.2 $\pm$ 0.7)  & \cite{1,Bunakov1981} \\
 4.4 $\pm$ 0.6 & \cite{2,BENKOULA1977}  & 11.2 $\pm$ 2.6  & \cite{1}  \\
  -(4.1 $\pm$ 0.8) & \cite{BUNAKOV1997} &  6.9 $\pm$ 0.8  & \cite{1}  \\ 
  8.9 $\pm$ 1.5 & \cite{Sushkov1982}   & 9.8 $\pm$ 4.1 & \cite{Olkhovsky1983}   \\
   8.5 $\pm$ 1.5  & \cite{BUNAKOV1983}  &  16.0 $\pm$ 2.1 & \cite{Bunakov1981}   \\
  & &  -(9.78 $\pm$ 4.08) & \cite{Bunakov1981}  \\
  \hline 
  
 $d\phi_{th}/dz$ ($10^{-6}$) (rad/cm) &  & $P_{p}$ ($10^{-3}$) &  \\ 
 \hline
 -36.7 $\pm$ 2.7 & \cite{1,2} & -(4.5 $\pm$ 1.3)  & \cite{2,Olkhovsky1983} \\
  -38.6 $\pm$ 6.8 & \cite{HECKEL1982} & 7.7 $\pm$ 1.3 & \cite{1,Smith2001} \\
    & & 7.9 $\pm$ 0.4 & \cite{Smith2001} \\ 
  \hline
  \hline
\end{tabular}
 \label{tab:P-odd_Ref}
\end{table}

\newpage
\section{Conclusions}

Our analysis shows
  that a resonance description of neutron radiative capture \cite{Flambaum:1983ek}   can describe all of the data taken so far at the 1.33 eV resonance in $^{117}$Sn, which is one of the most experimentally studied p-wave compound nuclear  resonance  to date.
  We observed an essential improvement in the fit of the experimental data without additional free parameters when we use the three-level approximation with destructive interference. This shows that the s-wave resonance at 38.8 eV is important  to describe the P-even effects at the vicinity of  p-wave resonance  in $^{117}$Sn, which are  dominant by the  mixing with the subthreshold s-wave resonance. It should be noted that
  to improve the accuracy of the measured resonance parameters the further analysis of new observables involving  gamma circular polarization and the $^{117}$Sn nuclear polarization is highly desirable. However, even at this stage our results show the assurance in understanding spectroscopical parameters which  are required for future experiments for the search of TRIV in neutron scattering on $^{117}$Sn target. The reasonably large  value  of the parameter $\kappa$  makes $^{117}$Sn a good candidate for a TRIV test.

We plan to apply a similar analysis for other candidates for the targets for TRIV experiments such as  $^{139}$La, $^{81}$Br, $^{131}$Xe, when more experimental data will be available.
%%%%%%%%%%%%%%%%%%%%%%%%%%%%%%%%%%%%%%%%%%%%%%%%%%%%%%%%%%%%%%%%%%%%%%

\section{Acknowledgment}
 L. Barr\'on-Palos acknowledges the support from PAPIIT-UNAM grant IN109120. \\ L. E. Char\'on-Garc\'ia acknowledges the support from CONACYT (Consejo Nacional de Ciencia y Tecnología). \\
V. Gudkov acknowledges support from U.S. Department of Energy Office of Science, Office of Nuclear Physics program under Award No. DE-SC0020687. \\
W. M. Snow acknowledges support from US National Science Foundation grants PHY-1913789 and PHY-2209481 and the Indiana University Center for Spacetime Symmetries. 

\nocite{*}

\bibliography{apssamp}% Produces the bibliography via BibTeX.

\end{document}